\documentclass{article}

\usepackage[a4paper, total={7in, 9.5in}]{geometry}
\usepackage{graphicx}
\usepackage{amsmath,amssymb}
\makeatletter
\newcommand{\biggg}{\bBigg@{3.5}}
\newcommand{\Biggg}{\bBigg@{4.5}}
\makeatother

\title{Transverse impulse response of finite and infinite Rayleigh beams}

\author{Saranya Biswas \and Anindya Chatterjee} 

\date{Indian Institute of Technology, Kanpur 208016, India.}

\begin{document}
	
\maketitle


\begin{abstract}
	The short-time impulse response of Euler-Bernoulli beams has been studied earlier, and shows unphysical transmission of high frequencies at unbounded speeds. The Rayleigh beam model adopts Euler-Bernoulli kinematics but adds sectionwise rotary inertia, leading to bounded wave speeds and physically better predictions of dynamic responses. Here, we study an undamped Rayleigh beam subjected to a concentrated transverse impulse applied at a point far from the boundaries. For a simply supported finite beam struck at its midpoint, using the modal expansion of the impulse response, a short-time asymptotic series for the response of the struck point is derived. The infinite beam impulse response is obtained as a spectral integral in two ways. The spectral integral is evaluated exactly in terms of modified Bessel functions via Laplace transform calculations. The corresponding short- and long-time asymptotics of the exact integral are consistent with the power series developed earlier and the Euler-Bernoulli beam, respectively. Direct numerical evaluation of the integral with a pre-specified error bound and numerical contour integration is also presented for verification. An ad hoc approximation is given which matches the response well for all times. The relationship between finite and infinite beam responses is examined in some detail. There are three consequences of finite length. First, the coefficients in the power series expansion contain components that decay exponentially with increasing length of the beam. Second, a smooth deviation from the infinite beam response is observed, consistent with the greatest group velocity in the beam, wherein a small precursor is detected which is consistent with broader theory. Third, a slope discontinuity is observed, corresponding to the high-frequency limit of group velocity. The full time series plot from modal expansion satisfactorily shows both the arrival and passage of these high frequency components. This paper opens the door to impact studies that are envisaged in future work.
\end{abstract}

\section{Introduction}
\label{intro}

Beams are widely used structural elements in civil and mechanical engineering, and their dynamic behaviour can be described at different levels of modelling approximation. The simplest of these is the Euler-Bernoulli model, which describes flexural motion by neglecting both the rotary inertia of the cross section and the transverse shear deformation. These assumptions are appropriate for sufficiently slender beams and for motions dominated by relatively long wavelengths. The Rayleigh beam model extends the Euler-Bernoulli beam theory by retaining rotary inertia while continuing to neglect transverse shear deformation. It thus incorporates an important inertial effect without introducing the additional field variable required in the Timoshenko theory. Details on the classical beam theories and their dynamic behaviour can be found in \cite{rayleigh1894theory,timoshenko1930strength,meirovitch1967analytical,graff2012wave}.

The inclusion of rotary inertia significantly modifies the high-frequency behaviour of the beam. In particular, unlike the Euler-Bernoulli beam, where wave speed increases without bound as the wavelength decreases, the wave speed in a Rayleigh beam approaches a finite limiting value at high frequencies.

The impulse response of the Euler-Bernoulli and Rayleigh beam models has been considered for different forms of impulse excitation in the literature. Chatterjee \cite{chatterjee2004short} studied an undamped simply supported Euler-Bernoulli beam subjected to a concentrated transverse impulse applied away from the boundaries. The analysis there considered the modal solution together with the corresponding infinite beam response and obtained detailed short-time asymptotic descriptions. Goswami {\it et al.} \cite{goswami2023short}, on the other hand, considered a concentrated angular impulse on a simply supported Rayleigh beam and derived its short-time response. Thus, the transverse impulse and angular impulse problems have been treated for different beam models, leaving the corresponding transverse impulse problem for the Rayleigh beam as a remaining fundamental case to consider. Note that, the impulse response of the infinite Euler-Bernoulli beam is expected to have some similarities with that of the Rayleigh beam, especially at the struck point and for large times. In addition, the asymptotic development of Goswami's paper will be adapted in our paper to obtain a portion of our results. In this way, \cite{chatterjee2004short,goswami2023short} together lead to the present paper.

In the present work, we study the response of finite and infinite undamped Rayleigh beams subjected to a concentrated transverse impulse applied at a point sufficiently far from the boundaries. In particular, for the finite case, we consider a simply supported beam struck at its midpoint. For a simply supported finite beam, an exact modal solution of the impulse response is available for all spatial points along the beam, and we obtain its short-time asymptotic expansion at the point of application of the impulse. The corresponding infinite beam response is derived, again for all spatial points and in the form of a spectral integral, both from the continuum limit of the finite beam modal solution and independently by Fourier transformation. At the point of application of the impulse, this integral is evaluated analytically in the Laplace-domain, subsequently yielding an exact time-domain expression of the infinite beam impulse response in terms of modified Bessel functions of the first kind. The corresponding short- and long-time asymptotic expansions are also presented. The integral expression for the infinite beam impulse response at the point of application of the impulse requires some care, and is numerically evaluated for illustration and comparison with the exact analytical solution. A simple approximation of the exact infinite beam impulse response at the point of application is also presented and compares well with the exact solution. The relation between the finite and the infinite beam responses is then examined by studying the finite-length and boundary-reflection effects. The finite-length correction is obtained analytically using the method of images. The onset of the boundary-reflection effect is consistent with the group velocity of the travelling waves in the Rayleigh beam. 

The rest of this paper is organized as follows. Section \ref{rayleigh} introduces the Rayleigh beam model and governing equation in dimensional and nondimensional forms. The exact finite beam impulse response is written as a discrete modal solution in Section \ref{modal}, and its short-time asymptotic behaviour is derived in Section \ref{emsum}. Section \ref{contspecanal} develops the exact impulse response of the infinite Rayleigh beam through the continuum limit of the finite beam modal solution and, independently, through a Fourier-transform approach. Its short- and long-time asymptotic forms are then obtained in Section \ref{contspecasym}. Sections \ref{approxinf} and \ref{sqrtapprox} consider two computational treatments of the infinite beam impulse response: a direct split-domain numerical evaluation of the spectral integral, and a simple ad hoc approximation of the impulse response, respectively. Their accuracies are compared in Section \ref{comparison}. In Section \ref{infval}, the finite and the infinite beam impulse responses are studied in some detail, and their difference is analytically examined and numerically verified. Concluding remarks are presented in Section \ref{conc}.


\section{Rayleigh beam model} \label{rayleigh}

\subsection{Governing equation}
\begin{figure}[htbp]
	\centering
	\includegraphics[width=0.58\textwidth]{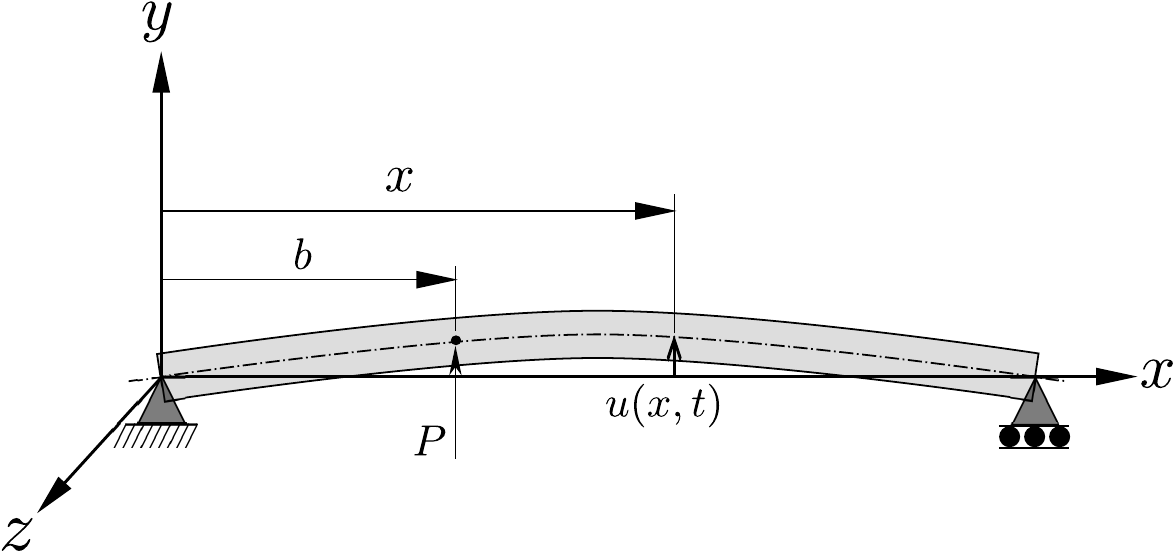}
	\caption{Schematic diagram of a simply supported beam with concentrated transverse impulse load.}
	\label{figschem}
\end{figure}

We consider a simply supported undamped Rayleigh beam of length $L$, as shown in Fig.\ \ref{figschem}, having flexural rigidity $EI$, mass per unit length $\rho A$, and rotary inertia of the cross section per unit length $\rho I$. The beam is subjected to a concentrated transverse impulse of magnitude $P$, applied at time $t=0$ and at a location $x=b$. The transverse displacement of the beam is denoted by $u(x,t)$. The governing equation of motion is
\begin{equation}
	EI\frac {\partial^4{u}}{\partial x^4}
	+\rho A \frac{\partial^2 u}{\partial t^2}
	- \rho I \frac{\partial^4 u}{\partial x^2 \partial t^2}
	=P \delta (x-b) \delta (t),
	\label{eq1}
\end{equation}
where $t$ is time, and the beam starts from rest, {\it i.e.}, zero initial conditions.

The inclusion of the rotary inertia term distinguishes the Rayleigh beam from the usual Euler-Bernoulli beam model. The rotary inertia is important in short-time or high-frequency, and small wavelength regimes.

\subsection{Nondimensionalization}
\label{dim}
Consider a long Rayleigh beam governed by Eq.\ \eqref{eq1}. The transverse displacement $u$ depends only on the quantities $EI$, $\rho A$, $\rho I$, $P$, and $t$, and is independent of the beam length $L$, assuming there is no reflected wave back to the point of observation. This assumption is exact for an infinitely long beam at all times, and also provides an excellent approximation for a sufficiently long finite beam over short times. Using Buckingham's $\Pi$-theorem of dimensional analysis, we can write
\begin{equation}
	u \sqrt{\dfrac{A}{I}}
	=
	\psi\!\left(
	\dfrac{P}{\sqrt{EA\,\rho I}},
	t\sqrt{\dfrac{EA}{\rho I}}
	\right),
	\label{eqdim1}
\end{equation}
where $\psi$ is a dimensionless function, and the three nondimensional terms are chosen to individually contain $u$, $P$ and $t$, respectively. Since the governing equation is linear, the response $u$ is proportional to the impulse magnitude $P$. Thus, recasting Eq.\ \eqref{eqdim1}, we write
\begin{equation}
	\bar u= \bar P \phi\,(\tau),\qquad \text{where} \  \ \bar u = u \sqrt{\dfrac{A}{I}},\quad  \bar P = \dfrac{P}{\sqrt{EA\,\rho I}}, \quad \tau = t\sqrt{\dfrac{EA}{\rho I}}.
	\label{eqdim2}
\end{equation}
In Eq.\ \eqref{eqdim2}, $\phi$ is a dimensionless function, $\bar u$ is the nondimensional displacement, $\bar P$ is the nondimensional impulse, and $\tau$ is the nondimensional time. Guided by Eq.\ \eqref{eqdim2}, we further choose the following nondimensional terms
\begin{equation}
	\bar x = x \sqrt{\dfrac{A}{I}}, \quad \bar b = b\sqrt{\dfrac{A}{I}}, \quad \bar L = L\sqrt{\dfrac{A}{I}}, \quad \bar v=v\sqrt{\dfrac{\rho}{E}},
	\label{eq1nondscale}
\end{equation}
where $v = \partial u / \partial t$ and $\bar v = \partial \bar u / \partial \tau$. The corresponding nondimensional governing equation is
\begin{equation}
	\dfrac {\partial^4{\bar u}}{\partial \bar x^4}
	+\dfrac{\partial^2 \bar u}{\partial \tau^2}
	- \dfrac{\partial^4 \bar u}{\partial \bar x^2 \partial \tau^2}
	=\bar P \delta (\bar x- \bar b) \delta (\tau).
	\label{eq1nond}
\end{equation}

\section{Exact impulse response of a finite beam}
\label{secfinite}

\subsection{Modal solution}
\label{modal}

The impulse response of Eq.\ \eqref{eq1nond} for a finite simply supported Rayleigh beam is first written by expanding the displacement in the beam's normal modes as
\begin{equation}
	\bar u(\bar x,\tau)=\sum_{k=1}^{\infty}a_k(\tau)\sin\left(\dfrac{k\pi \bar x}{\bar L}\right),
	\label{eqmod1}
\end{equation}
where $a_k(0)=0$. Let $\bar p_k =\dfrac{k\pi}{\bar L}$. Substituting Eq.\ \eqref{eqmod1} into Eq.\ \eqref{eq1}, multiplying by $\sin (p_n x)$, $n\in \mathbb{Z}^+$, and integrating over the beam length using the orthogonality of the mode shapes gives
\begin{equation}
	a_n''+\omega_n^2a_n=
	\dfrac{2 \bar P\sin(\bar p_n \bar b)}
	{\bar L(1+ \bar p_n^2)}
	\delta(\tau),  \qquad \omega_n^2=
	\dfrac{\bar p_n^4}
	{1+ \bar p_n^2},
	\label{eqmod2}
\end{equation}
where the primes denote derivatives with respect to the nondimensional time $\tau$. For a beam initially at rest, integration of Eq.\ \eqref{eqmod2} across $\tau=0$ gives
\begin{equation}
	a_n'(0^-)=0, \qquad
	a_n'(0^+)=
	\dfrac{2\bar P\sin(\bar p_n \bar b)}
	{\bar L(1+\bar p_n^2)}, \qquad \text{with} \quad a_n(0^-)=a_n(0^+)=0.
	\label{eqmod3}
\end{equation}
Hence, the solution of Eq.\ \eqref{eqmod2} is
\begin{equation}
	a_n(\tau)=
	\dfrac{2\bar P\sin(\bar p_n \bar b)}
	{\bar L(1+\bar p_n^2)\,\omega_n}
	\sin(\omega_n\tau).
	\label{eqmod4}
\end{equation}
The modal coordinates in Eq.\ \eqref{eqmod4} become
\begin{equation}
	a_n(\tau) =
	\dfrac{2\bar P}
	{\bar L}
	\dfrac{\sin(\bar p_n\bar b)}
	{\bar p_n^2\sqrt{1+\bar p_n^2}}
	\sin\left(
	\dfrac{\bar p_n^2}
	{\sqrt{1+\bar p_n^2}}
	\,\tau
	\right).
	\label{eqmod6}
\end{equation}
Substituting Eq.\ \eqref{eqmod6} into Eq.\ \eqref{eqmod1} gives
\begin{equation}
	\bar u(\bar x,\tau)=
	\dfrac{2 \bar P}
	{\bar L}
	\sum_{n=1}^{\infty}
	\dfrac{\sin(\bar p_n\bar b)\sin(\bar p_n\bar x)}
	{\bar p_n^2\sqrt{1+\bar p_n^2}}
	\sin\left(
	\dfrac{\bar p_n^2}
	{\sqrt{1+\bar p_n^2}}
	\,\tau
	\right).
	\label{eqmod7}
\end{equation}
Equation \eqref{eqmod7} gives the exact impulse response of the finite Rayleigh beam for an impulse applied at $\bar x = \bar b$, with the response evaluated at any observation point $0 \leq x \leq \bar L$ along the beam.

We now consider the impulse response at the point of application of the load, {\it i.e.}, $\bar x = \bar b$. For a sufficiently long beam at short times, the exact location of $b$ is immaterial provided it is sufficiently far from the supports. Choosing $\bar b=\bar L/2$, eliminates the even modes, and the impulse response becomes
\begin{equation}
	\bar u\left(\dfrac{\bar L}{2},\tau\right)
	=
	\dfrac{2 \bar P}
	{\bar L}
	\sum_{n=1}^{\infty}
	\dfrac{1}
	{(2n-1)^2q^2\sqrt{1+(2n-1)^2q^2}}
	\sin\left(
	\dfrac{(2n-1)^2q^2}
	{\sqrt{1+(2n-1)^2q^2}}
	\,\tau
	\right), \quad q=\dfrac{\pi}{\bar L}.
	\label{eqmod8}
\end{equation}
Equation \eqref{eqmod8} is exact. However, its short-time behaviour is not readily apparent and will be studied in the next section through an asymptotic expansion.
\begin{figure}[htbp]
	\centering
	\includegraphics[width=0.5\textwidth]{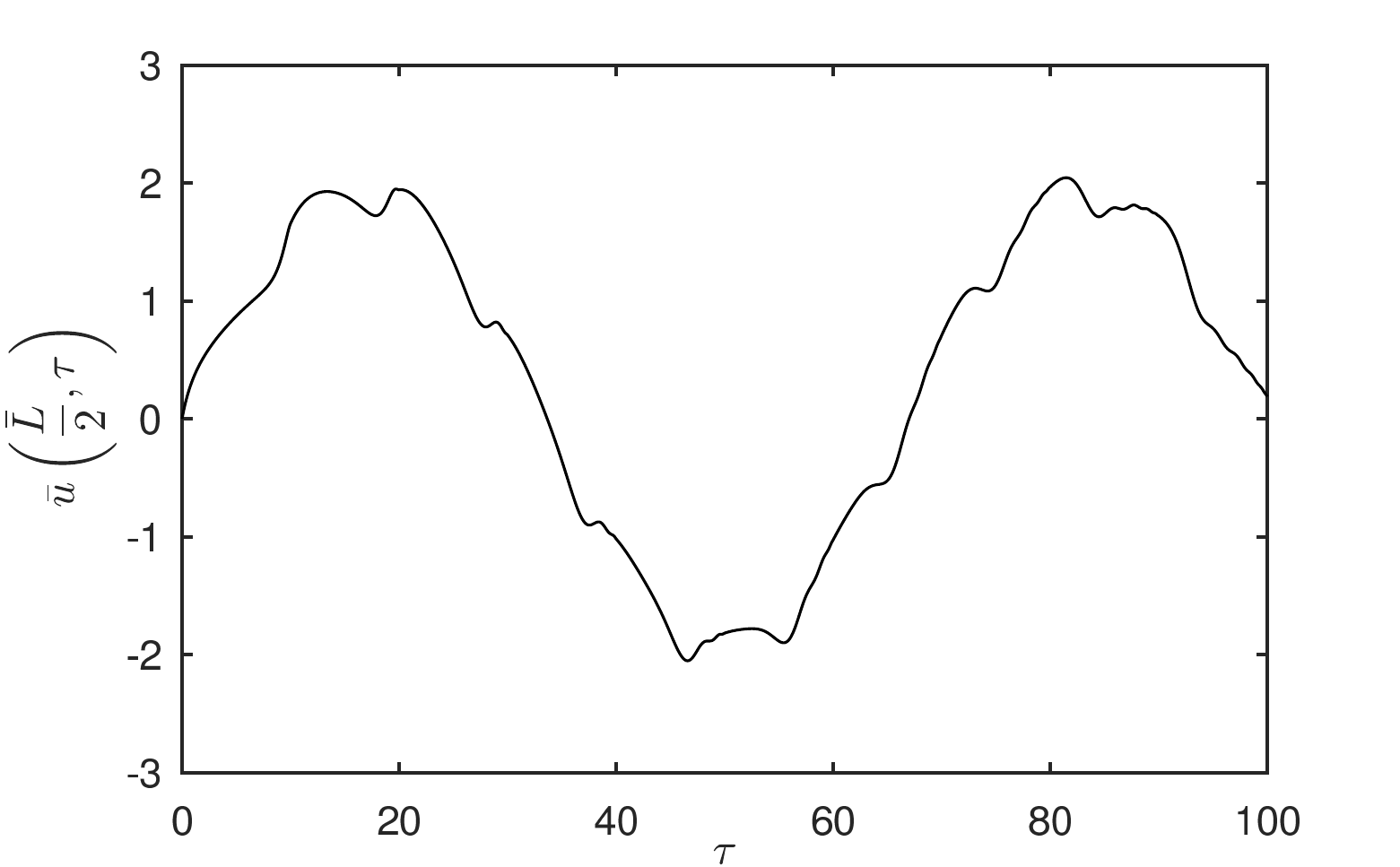}
	\caption{Impulse response $u\!\left(\dfrac{\bar L}{2},\tau\right)$ from Eq.\ \eqref{eqmod8}, for $\bar L=10$ and $\bar P=1$. $10^3$ modes are retained. The time step is $10^{-2}$.}
	\label{series}
\end{figure}

Figure \ref{series} shows the impulse response at $\bar L/2$ obtained from Eq.\ \eqref{eqmod8} for $\bar L=10$ and $\bar P=1$. The response is evaluated up to $\tau=100$ with a time step of $10^{-2}$, and the series is truncated at $10^3$ terms, which is sufficient for graphing accuracy.

\begin{figure}[htbp]
	\centering
	\includegraphics[width=\textwidth]{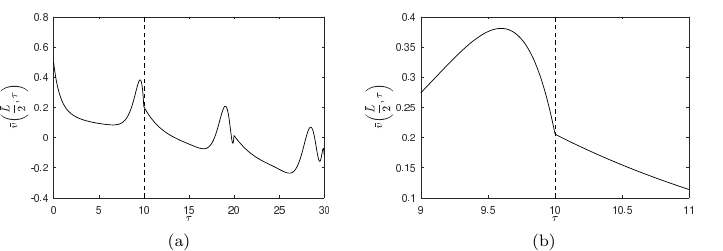}	
	\caption{For $\bar L=10$ and $\bar P=1$, (a) velocity response at the point of application of the impulse, upto $\tau = 30$, $10^3$ modes are retained, time step is $10^{-2}$, (b) enlarged view near $\tau=10$.}
	\label{fig:velocity}
\end{figure}

Differentiating Eq.\ \eqref{eqmod8} with respect to $\tau$ gives the corresponding velocity,
\begin{equation}
	\bar v\left(\dfrac{\bar L}{2},\tau\right)
	=\dfrac{2\bar P}
	{\bar L}\sum_{n=1}^{\infty}\dfrac{1}{(2 n-1)^2 q^2+1}\cos \left(\dfrac{(2 n-1)^2 q^2 \tau }{\sqrt{(2 n-1)^2 q^2+1}}\right),
	\label{eqvel}
\end{equation}
which is useful for identifying the arrival of wave reflections from the beam's endpoints.

Figure \ref{fig:velocity}(a) shows the velocity response at the point of application of the impulse. Unlike the displacement, the velocity exhibits a slope discontinuity at $\tau=10$, as seen in Fig.\ \ref{fig:velocity}(b). This slope discontinuity corresponds to the arrival of reflections at very high frequencies from the nearest beam boundary. Prior to this instant, the response is almost identical to that of an effectively infinite beam. We say `almost identical' because there are some finite-length effects and boundary-reflection effects before $\tau=10$. Those deviations will be studied in Section \ref{infval}.

\subsection{Short-time asymptotics of the impulse response of a finite beam} \label{emsum}
The short-time asymptotic expansion of the solution represented by Eq.\ \eqref{eqmod8} is derived following the methodology of Goswami {\it et al.} \cite{goswami2023short}, adapted to the present problem. Since the details are lengthy, the complete derivation is presented in Appendix \ref{appn_asymcal}. The resulting asymptotic expansion of Eq.\ \eqref{eqmod8} for $\tau\ll1$ is
\begin{equation}
	\begin{aligned}
		\bar u\left(\dfrac{\bar L}{2},\tau\right)_{asymp}=& \bar P \Bigg[ \dfrac{\tau}{2}   \tanh \left(\dfrac{\bar L}{2}\right) - \dfrac{\tau^2}{4} + \left\{\dfrac{1}{8} \tanh \left(\dfrac{\bar L}{2}\right)+\dfrac{1}{48} \bar L \text{sech}^2\left(\dfrac{\bar L}{2}\right)\right\}\tau ^3 -\dfrac{5 \tau ^4}{96} \\ & +\left(-\dfrac{1}{480} \bar L^2 \sinh ^4\left(\dfrac{\bar L}{2}\right) \text{csch}^3(\bar L)+\dfrac{7}{384} \tanh \left(\dfrac{\bar L}{2}\right)+\dfrac{13 \bar L \text{sech}^2\left(\dfrac{\bar L}{2}\right)}{3840}\right)\tau ^5  + \mathcal{O}(\tau^6) \Bigg].
	\end{aligned}
	\label{eq28_full}
\end{equation}
Equation \eqref{eq28_full} gives the short-time asymptotic expansion of the impulse response at the point of application of the concentrated impulse, $\bar x=\bar L/2$. It is seen that the coefficients in the power series depend on $\bar L$, but rapidly attain limiting values for large $\bar L$.

The corresponding short-time asymptotic expansion for an infinitely long Rayleigh beam follows by taking the limit $\bar L\rightarrow\infty$, for which $\tanh\left(\dfrac{\bar L}{2}\right) \rightarrow 1$, $\bar L \text{sech}^2\left(\dfrac{\bar L}{2}\right) \rightarrow 0$, and $\bar L^2 \sinh ^4\left(\dfrac{\bar L}{2}\right) \text{csch}^3(\bar L) \rightarrow 0$, reducing Eq.\ \eqref{eq28_full} to
\begin{equation}
	\bar u_\infty(0,\tau)_{asymp}=\bar P \left(\dfrac{\tau }{2}-\dfrac{\tau ^2}{4}+\dfrac{\tau ^3}{8}-\dfrac{5 \tau ^4}{96}+\dfrac{7 \tau ^5}{384} + \mathcal{O}(\tau^6)\right),
	\label{eq29_full}
\end{equation}
where we shift the coordinate $\bar x$ to let $\bar x =0$ be the point of application of the impulse as well as the point of observation for an infinite beam, and use a subscript as in $\bar u_\infty (0,\tau)$.

\begin{figure}[htbp]
	\centering
	\includegraphics[width=\textwidth]{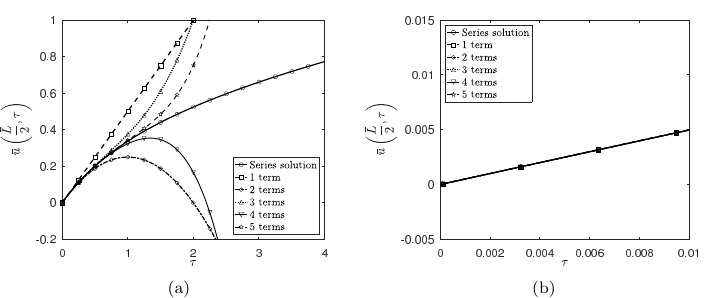}
	\caption{(a) Comparison of the full solution of Eq.\ \eqref{eqmod8} with successive truncations of the short-time asymptotic expansion of Eq.\ \eqref{eq28_full}, (b) enlarged view near $\tau=0$, for $\bar L=10$ and $\bar P=1$.}
	\label{serasym}
\end{figure}
\begin{figure}[htbp]
	\centering
	\includegraphics[width=0.46\textwidth]{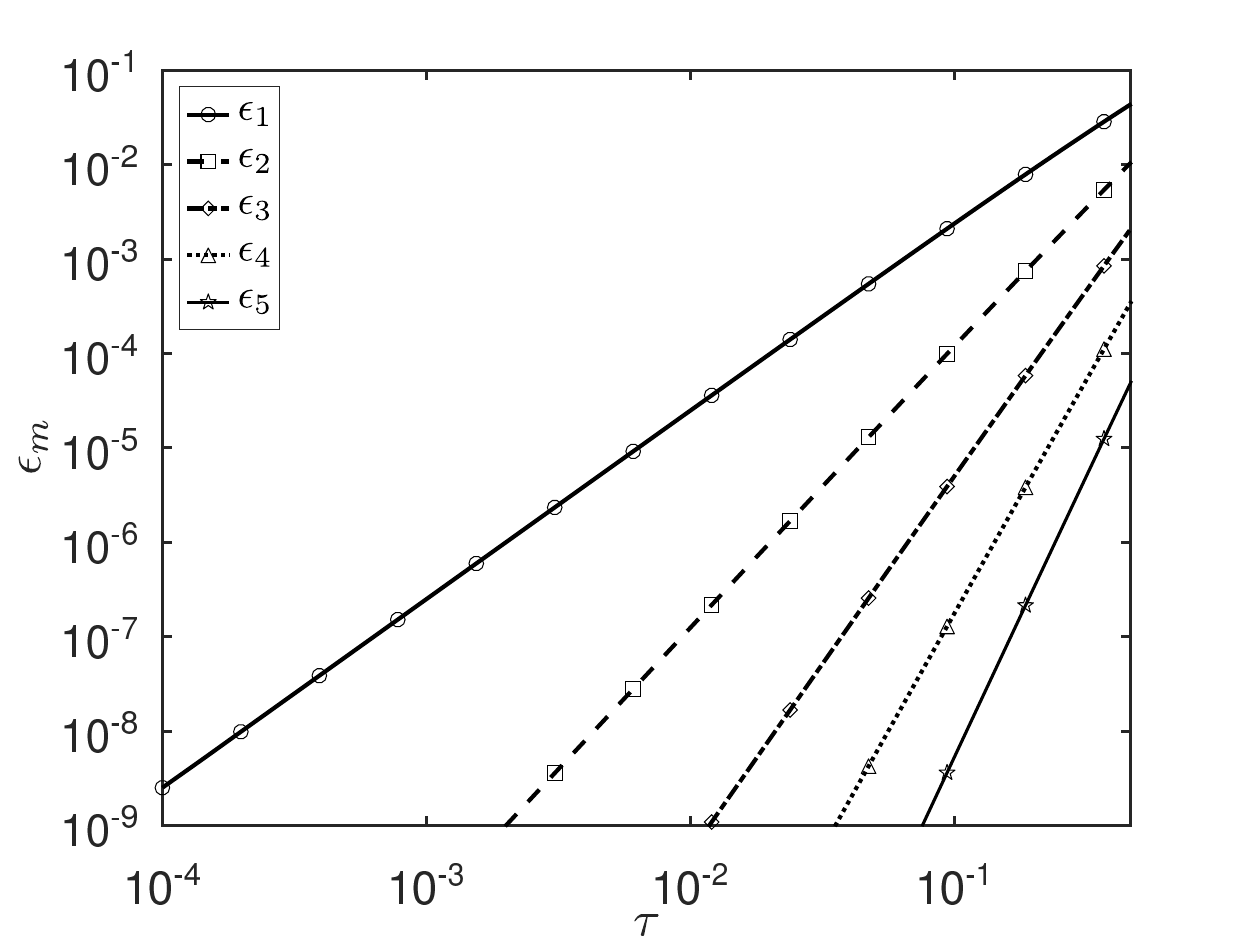}
	\caption{Absolute error between the full modal solution and successive truncations of the short-time asymptotic expansion. The observed power-law behaviour agrees with the predicted order $\epsilon_m=\mathcal{O}(\tau^{m+1})$.}
	\label{err2}
\end{figure}

The response is proportional to the applied impulse $P$ and depends only on the non-dimensional time $\tau$, consistent with Eq.\ \eqref{eqdim2}. It is also interesting to note that the short-time impulse response of a Rayleigh beam has a leading term proportional to $\tau$, unlike the Euler-Bernoulli beam where it is proportional to $\sqrt{\tau}$ (see \cite{chatterjee2004short}).

The asymptotic expansion Eq.\ \eqref{eq28_full} can be compared with the modal solution in Eq.\ \eqref{eqmod8}. Figure \ref{serasym} compares successive truncations of Eq.\ \eqref{eq28_full} with the modal solution for $\bar L=10$ and $\bar P=1$. The asymptotic approximations approach the modal solution as $\tau\rightarrow0$, and retaining more terms yields a better match for small $\tau$.

For a more careful verification, let $\bar u_m\left(\dfrac{\bar L}{2},\tau\right)_{asymp}$ denote the asymptotic expansion truncated after the term of order $\tau^m$. We define the absolute error as
\begin{equation}
	\epsilon_m=
	\left|
	\bar u\left(\dfrac{\bar L}{2},\tau\right)-\bar u_m\left(\dfrac{\bar L}{2},\tau\right)_{asymp}
	\right|.
	\label{eqerr1}
\end{equation}
Truncation after $\tau^m$ requires $\epsilon_m=\mathcal{O}(\tau^{m+1})$ as $\tau\rightarrow0$. Figure \ref{err2} shows the corresponding errors on logarithmic axes. The observed power-law behaviours agree with theory in each case.


\section{Exact impulse response of an infinite beam}
\label{contspec}

\subsection{Analytical solution}
\label{contspecanal}
For an infinitely long beam, the discrete spectrum of the finite beam
becomes continuous. The impulse response of the infinite beam can
therefore be obtained by taking the continuum limit of the finite beam
modal solution. The same continuous spectral representation can also
be obtained independently by applying the spatial Fourier transform to the governing equation. We first obtain the displacement response at an arbitrary spatial point $\bar x$ along the infinite beam. The response at the point of application of the impulse is then considered as a special case with $\bar x =0$.

Recall the finite beam modal solution in Eq.\ \eqref{eqmod7},
\begin{equation}
	\bar u(\bar x,\tau)=
	\dfrac{2 \bar P}
	{\bar L}
	\sum_{n=1}^{\infty}
	\dfrac{\sin(\bar p_n\bar b)\sin(\bar p_n\bar x)}
	{\bar p_n^2\sqrt{1+\bar p_n^2}}
	\sin\left(
	\dfrac{\bar p_n^2}
	{\sqrt{1+\bar p_n^2}}
	\,\tau
	\right), \qquad \bar p_n=\dfrac{n\pi}{\bar L}.
	\label{eqinf1} 
\end{equation}
As $\bar L\rightarrow\infty$, the spacing between successive
wavenumbers, $\Delta \bar p=\dfrac{\pi}{\bar L}$, tends to zero. Hence, the summation in Eq.\ \eqref{eqinf1} becomes a
continuous spectral integral,
\begin{equation}
	\bar u_\infty(\bar x,\tau)
	=
	\dfrac{2\bar P}{\pi}
	\int_0^\infty
	\dfrac{
		\sin( \kappa \bar b)\sin( \kappa \bar x)}
	{ \kappa ^2\sqrt{1+ \kappa ^2}}
	\sin\left(
	\dfrac{ \kappa ^2}{\sqrt{1+ \kappa ^2}}\tau
	\right)
	\,{\rm d} \kappa ,
	\label{eqinf2}
\end{equation}
where $ \kappa $ is a continuous wavenumber.

A shift in coordinates is now made. The point of application of the impulse is now chosen as the origin. Let the spatial coordinate measured from this point be denoted by $\bar x$, so that the corresponding coordinate in the finite beam is $\bar b+\bar x$. Using $\sin( \kappa \bar b)\sin\left[ \kappa (\bar b+\bar x)\right]
=
\dfrac{1}{2}
\left[
\cos( \kappa \bar x)
-
\cos\left( \kappa (2\bar b+\bar x)\right)
\right]$, Eq.\ \eqref{eqinf2} can be rewritten as
\begin{equation}
	\bar u_\infty(\bar x,\tau)
	=
	\dfrac{\bar P}{\pi}
	\int_0^\infty
	\dfrac{
		\cos( \kappa \bar x)
		-
		\cos\left( \kappa (2\bar b+\bar x)\right)}
	{ \kappa ^2\sqrt{1+ \kappa ^2}} \sin\left(
	\dfrac{ \kappa ^2}{\sqrt{1+ \kappa ^2}}\tau
	\right)
	\,{\rm d} \kappa .
	\label{eqinf3}
\end{equation}
In the infinite beam limit ($\bar b \to \infty$), the second term in
Eq.\ \eqref{eqinf3} involves the rapidly oscillating factor
$\cos\left( \kappa (2\bar b+\bar x)\right)$, while the rest of the integrand remains absolutely integrable. By the Riemann-Lebesgue lemma \cite{serov2017riemann}, this term vanishes as $\bar b \to \infty$. Equation \eqref{eqinf3} thus reduces to
\begin{equation}
	\bar u_\infty(\bar x,\tau)
	=
	\dfrac{\bar P}{\pi}
	\int_0^\infty
	\dfrac{
		\sin\left(
		\dfrac{ \kappa ^2}{\sqrt{1+ \kappa ^2}}\tau
		\right)
		\cos( \kappa \bar x)}
	{ \kappa ^2\sqrt{1+ \kappa ^2}}
	\,{\rm d} \kappa .
	\label{eqinfx}
\end{equation}
The same result can also be obtained directly from the governing equation. Taking the spatial Fourier transform of Eq.\ \eqref{eq1nond} gives
\begin{equation}
	(1+ \kappa ^2)
	\dfrac{\partial^2\hat u}{\partial\tau^2}
	+ \kappa ^4\hat u
	=
	\bar P\delta(\tau),
	\label{eqsecn10a}
\end{equation}
where $\hat u( \kappa ,\tau)$ is the spatial Fourier transform of the
displacement. The solution of Eq.\ \eqref{eqsecn10a} is
\begin{equation}
	\hat u( \kappa ,\tau)
	=
	\dfrac{\bar P
		\sin\left(
		\dfrac{ \kappa ^2}{\sqrt{1+ \kappa ^2}}\tau
		\right)}
	{ \kappa ^2\sqrt{1+ \kappa ^2}}.
	\label{eq:general_fourier_solution}
\end{equation}
The inverse Fourier transform is
$ \displaystyle \bar u_\infty(\bar x,\tau)
=
\dfrac{1}{2\pi}
\int_{-\infty}^{\infty}
\hat u( \kappa ,\tau)\,e^{i \kappa \bar x}\,{\rm d} \kappa 
$, and since $\hat u( \kappa ,\tau)$ is an even function of $ \kappa $, $\displaystyle \bar u_\infty(\bar x,\tau)
=
\dfrac{1}{\pi}
\int_0^\infty
\hat u( \kappa ,\tau)\cos( \kappa \bar x)\,{\rm d} \kappa 
$.
Substitution of Eq.\ \eqref{eq:general_fourier_solution} gives
Eq.\ \eqref{eqinfx}. Thus, the limit of the infinite-length modal solution and the direct Fourier-transform approach agree.

We now consider the response at the point of application of the
impulse, namely, $\bar x=0$.
Equation \eqref{eqinfx} reduces to
\begin{equation}
	\bar u_\infty(0,\tau)
	=
	\dfrac{\bar P}{\pi}
	\int_0^\infty
	\dfrac{
		\sin\left(
		\dfrac{ \kappa ^2}{\sqrt{1+ \kappa ^2}}\tau
		\right)}
	{ \kappa ^2\sqrt{1+ \kappa ^2}}
	\,{\rm d} \kappa .
	\label{eqsecn10}
\end{equation}

We can evaluate the integral in Eq.\ \eqref{eqsecn10} analytically. Taking the Laplace transform with respect to $\tau$ and interchanging the order of integration gives
\begin{equation}
	\begin{aligned}
		\Tilde u(0,s)&=
		\mathcal{L}\left\{\dfrac{\bar P}{\pi }
		\int_0^\infty
		\dfrac{\sin\left(\dfrac{ \kappa ^2}{\sqrt{1+ \kappa ^2}}\tau\right)}
		{ \kappa ^2\sqrt{1+ \kappa ^2}}
		\,{\rm d} \kappa \right\}=\dfrac{\bar P}{\pi }
		\int_0^\infty \mathcal{L} \left \{
		\dfrac{\sin\left(\dfrac{ \kappa ^2}{\sqrt{1+ \kappa ^2}}\tau\right)}
		{ \kappa ^2\sqrt{1+ \kappa ^2}}\right \}
		\,{\rm d} \kappa  \\ &=\dfrac{\bar P}{\pi }
		\int_0^\infty
		\dfrac{1}
		{s^2(1+ \kappa ^2)+ \kappa ^4} {\rm d} \kappa ,
	\end{aligned}
	\label{eqsecn13}
\end{equation}
where $s$ is the Laplace variable. The resulting integral can be
evaluated in closed form, giving
\begin{equation}
	\Tilde u_\infty(0,s)
	=
	\dfrac{\bar P}
	{2s\sqrt{s(s+2)}}.
	\label{eqsecn14}
\end{equation}
The inverse Laplace transform of Eq.\ \eqref{eqsecn14} is
\begin{equation}
	\bar u_\infty(0,\tau)
	=
	\dfrac{\bar P}{2}
	\tau e^{-\tau}
	\left[
	I_0(\tau)+I_1(\tau)
	\right],
	\label{eqexactinf}
\end{equation}
where $I_n(\cdot)$ denotes a modified Bessel function of the first kind. This inverse can be obtained from symbolic mathematics packages like Mathematica. Equation \eqref{eqexactinf} gives the exact impulse response of
the infinite Rayleigh beam at the point of application of the impulse
at any arbitrary time.

The velocity response at an arbitrary spatial point is obtained
by differentiating Eq.\ \eqref{eqinfx} with respect to $\tau$, which
gives
\begin{equation}
	\bar v_\infty(\bar x,\tau)
	=
	\dfrac{\partial\bar u_\infty(\bar x,\tau)}
	{\partial\tau}
	=
	\dfrac{\bar P}{\pi}
	\int_0^\infty
	\dfrac{
		\cos\left(
		\dfrac{ \kappa ^2}{\sqrt{1+ \kappa ^2}}\tau
		\right)
		\cos( \kappa \bar x)}
	{1+ \kappa ^2}
	\,{\rm d} \kappa .
	\label{eqinfv}
\end{equation}
At the point of application of the impulse, $\bar x=0$, this becomes (integral form)
\begin{equation}
	\bar v_\infty(0,\tau)
	=
	\dfrac{\bar P}{\pi}
	\int_0^\infty
	\dfrac{
		\cos\left(
		\dfrac{ \kappa ^2}{\sqrt{1+ \kappa ^2}}\tau
		\right)}
	{1+ \kappa ^2}
	\,{\rm d} \kappa .
	\label{eqsecn10vel}
\end{equation}
Alternatively, differentiating the exact displacement response in
Eq.\ \eqref{eqexactinf} gives (closed form)
\begin{equation}
	\bar v_\infty(0,\tau)
	=
	\dfrac{\partial\bar u_\infty(0,\tau)}
	{\partial\tau}
	=
	\dfrac{\bar P}{2}
	e^{-\tau}I_0(\tau).
	\label{eqexactinfvel}
\end{equation}


\subsection{Short- and large-time asymptotics of the impulse response of an infinite beam}
\label{contspecasym}
Next, we derive the short-time and large-time asymptotic expansions of Eq.\ \eqref{eqexactinf}. The short-time expansion can be obtained by expanding Eq.\ \eqref{eqexactinf} about $\tau=0$. Equivalently, the same expansion can be obtained more conveniently to any desired order from the large-$s$ expansion of Eq.\ \eqref{eqsecn14}, followed by inverse Laplace transformation, yielding
\begin{equation}
	\bar u_\infty(0,\tau)_{asymp}
	=
	\bar P
	\left(
	\frac{\tau}{2}
	-\frac{\tau^2}{4}
	+\frac{\tau^3}{8}
	-\frac{5\tau^4}{96}
	+\frac{7\tau^5}{384}
	-\frac{7\tau^6}{1280}
	+\frac{11\tau^7}{7680}
	-\frac{143\tau^8}{430080}
	+\mathcal O(\tau^9)
	\right).
	\label{eqshortcont}
\end{equation}
The above matches Eq.\ \eqref{eq29_full}, verifying the short-time expansion obtained from the discrete-spectrum analysis.

Similarly for large time, it is simplest to take the small-$s$ expansion of Eq.\ \eqref{eqsecn14} and then compute the inverse Laplace transform to obtain
\begin{equation}
	\bar u_\infty(0,\tau)_{asymp,\infty}
	=
	\bar P
	\dfrac{\sqrt{\tau}}{\sqrt{2\pi}} \left[
	1
	-\dfrac{1}{8 \tau}
	-\dfrac{3}{128 \tau^2} -\dfrac{15}{1024 \tau^3} -\dfrac{525}{32768 \tau ^4} -\dfrac{6615}{262144 \tau ^5}
	+\mathcal O(\tau^{-6})
	\right],
	\qquad
	\tau\rightarrow\infty .
	\label{eqlongcont}
\end{equation}
The leading term agrees with the classical Euler-Bernoulli result \cite{chatterjee2004short}, showing that rotary inertia has a decaying effect for large times because the shape of the disturbance becomes broader with time.

It is emphasized that, unlike the Euler-Bernoulli beam, for the Rayleigh beam the short-time impulse response has a leading term proportional to $\tau$, and the large-time response has a leading term proportional to $\sqrt{\tau}$, leading to two different asymptotic regimes in time.

Interestingly, these asymptotic expansions can be computed from Eq.\ \eqref{eqsecn14}, and do not require the closed-form solution in Eq.\ \eqref{eqexactinf}.

\subsection{Numerical evaluation of the impulse response integral}
\label{approxinf}

We now evaluate the spectral integral in Eq.~\eqref{eqsecn10} numerically, without using the closed-form result in Eq.~\eqref{eqexactinf}. This is useful because that may be our only recourse when a closed-form impulse response is not available. We write the integral in Eq.~\eqref{eqsecn10} as
\begin{equation}
	I(\tau)=\int_0^\infty
	\frac{\sin\left(\tau g(\kappa)\right)}
	{\kappa^2\sqrt{1+\kappa^2}}
	\,{\rm d}\kappa,
	\qquad \text{where} \quad
	g(\kappa)=\frac{\kappa^2}{\sqrt{1+\kappa^2}},
	\label{eqdef}
\end{equation}
and seek a numerical value $I_{\rm num}(\tau)$ such that $|I(\tau)-I_{\rm num}(\tau)|\leq\varepsilon$, with $\varepsilon=10^{-10}$, for a long time range of $10^{-10}\leq \tau \leq 10^8$. Here $\varepsilon=10^{-10}$ is an arbitrarily chosen number in the range $10^{-16} \ll \varepsilon \ll 1$.

Consider the asymptotic behaviour of the integrand in Eq.~\eqref{eqdef} at the two ends of the range of integration. As $\kappa\to0$, the integrand is regular and approaches $\tau$; there is no difficulty here. As $\kappa\to\infty$, however, $g(\kappa)\sim\kappa$, so the integrand behaves as $\sin(\tau\kappa)/\kappa^3$ and oscillates increasingly fast as $\tau$ grows, making direct quadrature on the real axis progressively more expensive. This contrast motivates splitting the range of integration at a small point $\kappa=\sigma$ (to be chosen later). The integral over $[0,\sigma]$ can be evaluated directly on the real axis, while the integral over $[\sigma,\infty)$ can be treated differently to remove the oscillation. 

\begin{figure}[htbp]
	\centering
	\includegraphics[width=0.45\textwidth]{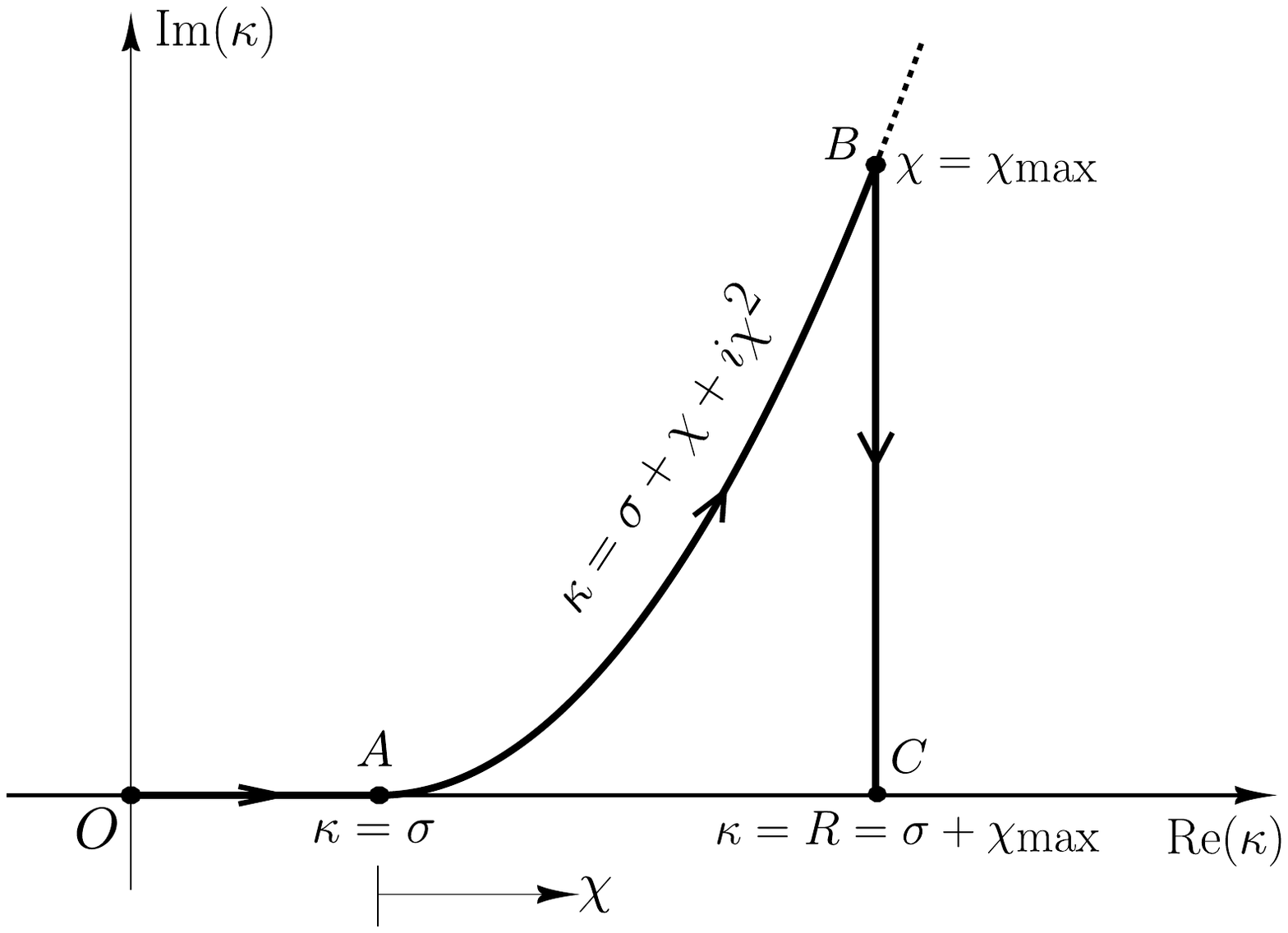}
	\caption{Schematic of the integration path. The integral is evaluated directly on the real-axis segment $OA$. From $A$, the path is deformed into the upper half plane along $AB$. The segment $BC$ formally completes the integral but makes zero contribution.}
	\label{fig:incontour}
\end{figure}

For the latter integral, we use
\begin{equation}
	\frac{\sin\left(\tau g(\kappa)\right)}
	{\kappa^2\sqrt{1+\kappa^2}}
	=
	\operatorname{Im}\left[
	\frac{e^{i\tau g(\kappa)}}
	{\kappa^2\sqrt{1+\kappa^2}}
	\right].
	\label{eqhdef}
\end{equation}
The exponential on the right decays if the path of integration is deformed into the upper half of the complex $\kappa$-plane. The integration path is shown in Fig.~\ref{fig:incontour}. The complete path starts at the origin and runs along the real axis up to a point $A$ at $\kappa=\sigma$; this segment is labelled $OA$, and along it the original real-valued integrand is used, as noted above. From $A$, the path is deformed into the upper half plane along a curved segment $AB$, upto a truncation point $B$ (to be chosen later), along which the complex form in Eq.~\eqref{eqhdef} is used instead. A vertical segment $BC$ completes the integral. As explained later, its contribution vanishes as its real coordinate is taken to infinity, so it is just omitted.

Our four main sources of numerical error are: quadrature error on the real axis, round-off error, truncation of the deformed path, and quadrature error on the deformed path. The total tolerance $\varepsilon$ is divided equally among these four, so each is allotted $\varepsilon/4$ for purposes of error estimation and control.

The value of $\sigma$ cannot be taken to be zero, since the complex form in Eq.~\eqref{eqhdef} behaves as $1/\kappa^2$ near $\kappa=0$, unlike the original integrand which is regular there. The value of $\sigma$ is set by floating-point round-off. Near $\kappa=\sigma$, the complex integrand is of order $1/\sigma^2$, so evaluating it introduces an absolute error of order $\varepsilon_{\rm mach}/\sigma^2$, where $\varepsilon_{\rm mach}\approx2.22\times10^{-16}$ is the machine precision in our GNU Octave based computations. This error acts over a stretch of the path of length of order $\sigma$, giving a net contribution to the integral of order $\varepsilon_{\rm mach}/\sigma$. Requiring this to stay within the allotted $\varepsilon/4$ gives
\begin{equation}
	\frac{\varepsilon}{4}=\frac{\varepsilon_{\rm mach}}{\sigma} \implies \sigma=\frac{4\,\varepsilon_{\rm mach}}{\varepsilon}.
	\label{eq:sigmamin}
\end{equation}

The integral over $OA$, from $0$ to $\sigma$, is evaluated directly using the original real-valued integrand. Even over this short interval, the phase $\tau g(\kappa)$ can change rapidly when $\tau$ is large, so the interval is divided into a number of subintervals $N_{OA}$ on path $OA$, based on the total phase change across $[0,\sigma]$,
\begin{equation}
	N_{OA}=\left\lceil \frac{\tau g(\sigma)}{\pi/4} \right\rceil,
	\label{eq:Ninner}
\end{equation}
where $\lceil \cdot \rceil$ is the ceiling function, and the phase step is chosen to be $\pi/4$. On each subinterval, the integral is evaluated using both 32-point and 64-point Gauss-Legendre quadrature. The 64-point value is retained. 

The sum of the absolute differences between the 32-point and 64-point values, over all subintervals, was numerically verified to be within the allotted tolerance $\varepsilon/4$. No further refinement of the subintervals was needed.

The remaining part of the integral, from $\sigma$ to $\infty$, is evaluated along the curved path $AB$, parameterized as
\begin{equation}
	\kappa(\chi)=\sigma+\chi+i\chi^2,
	\qquad
	\chi\geq0,
	\qquad
	\frac{{\rm d}\kappa}{{\rm d}\chi}=1+2i\chi,
	\label{eqparabola}
\end{equation}
so that the required contribution to $I(\tau)$ is the imaginary part of
\begin{equation}
	\int_0^\infty
	\frac{e^{i\tau g(\kappa(\chi))}}
	{\kappa(\chi)^2\sqrt{1+\kappa(\chi)^2}}
	\left(1+2i\chi\right)
	\,{\rm d}\chi.
	\label{eqparabolaexact}
\end{equation}
Along this path $g(\kappa(\chi))\sim\kappa(\chi)$ for large $\chi$, so $e^{i\tau g(\kappa(\chi))}\sim e^{i\tau(\sigma+\chi)}e^{-\tau\chi^2}$: the exponential now decays rapidly. The closing segment $BC$ contributes negligibly because the exponential does not grow in the right half plane while the remaining factor decays as $1/|\kappa|^3$; also, no singularity of the integrand is crossed in deforming the path from the real axis to $AB$, since the branch points at $\kappa=\pm i$ lie outside the region swept by the deformation.

The integral along $AB$ is truncated at a finite value $\chi=\chi_{\max}$ at $B$. A cutoff that works for all $\tau$ follows from the algebraic decay of the integral at large $\chi$, roughly as $\dfrac{3\sqrt{2}}{4\chi^4}$. After assigning the truncation error a tolerance of $\varepsilon/4$, this gives the cutoff
\begin{equation}
	\chi_{\max}=\left(\frac{3\sqrt{2}}{\varepsilon}\right)^{1/4}.
	\label{eqchia}
\end{equation}
This value is conservative and does not exploit the faster decay available at large $\tau$. Near the start of the path, the phase grows like $\tau\chi^2$, and this can be used to show that for large $\tau$, the integral along $AB$ decays approximately as $e^{-2\tau\chi^3}$. This gives a second, smaller cutoff
\begin{equation}
	\chi_{\max}(\tau)=\left[\frac{\ln(4/\varepsilon)}{2\tau}\right]^{1/3}.
	\label{eqchie}
\end{equation}
Equation~\eqref{eqchia} gives an estimate that is independent of $\tau$ and is always a valid cutoff, but it remains unnecessarily large for large $\tau$. Equation~\eqref{eqchie} decreases with $\tau$, but is only accurate once $\tau$ is large enough for the near-origin approximation behind it to hold. At each $\tau$, we therefore use whichever of the two is smaller: Eq.~\eqref{eqchia} at small and moderate $\tau$, and Eq.~\eqref{eqchie} once it drops below the value given by Eq.~\eqref{eqchia}. In this way the length of the integration path shrinks automatically as $\tau$ increases, instead of remaining fixed at an unnecessarily conservative value.

The integral along $AB$ up to $\chi_{\max}$ is again evaluated by Gauss-Legendre quadrature over a number of subintervals. Near the start of the path the phase behaves as $\tau\chi^2$, so it is convenient to measure distance along the path in the rescaled variable $\lambda=\sqrt{\tau}\,\chi$, in which the local oscillation no longer depends explicitly on $\tau$. Dividing the range uniformly in $\lambda$, using a fixed subinterval width $\Delta \lambda$, gives a number of subintervals on path $AB$ as
\begin{equation}
	N_{AB}=\left\lceil \frac{\sqrt{\tau}\,\chi_{\max}}{\Delta\lambda} \right\rceil,
	\qquad \Delta\lambda = 0.25.
	\label{eq:Nouter}
\end{equation}
As before, each subinterval is evaluated with both 32-point and 64-point Gauss-Legendre quadrature, the 64-point value is retained, and the summed absolute differences between the two rules over all subintervals is checked to remain within the allotted tolerance $\varepsilon/4$; this is again verified numerically, with no further refinement needed.

The numerical value of Eq.~\eqref{eqdef} is obtained by adding the real-axis contribution from $OA$ to the imaginary part of the contour contribution from $AB$. Since each of the four sources of error above is kept within $\varepsilon/4$, the total error satisfies
\begin{equation}
	|I(\tau)-I_{\rm num}(\tau)|
	\lesssim
	\frac{\varepsilon}{4}+\frac{\varepsilon}{4}+\frac{\varepsilon}{4}+\frac{\varepsilon}{4}
	=
	\varepsilon.
	\label{eqtotalerror1}
\end{equation}
The impulse response of the infinite Rayleigh beam then follows from $\bar u_{\rm num}(\tau)=\bar P\,I_{\rm num}(\tau)/\pi$. This numerical procedure uses only the continuum spectral integral in Eq.~\eqref{eqdef}; the closed-form solution in Eq.~\eqref{eqexactinf} is not used. We will examine the numerical accuracy of the integral after presenting a different approximation.


\subsection{Approximate ad hoc formulas}
\label{sqrtapprox}

We can construct a simple ad hoc approximation directly from the spectral integral in Eq.\ \eqref{eqsecn10}. This is potentially useful when the spectral integral is available but no exact time-domain solution is known analytically. The short- and large-time asymptotics are assumed available from the Laplace domain representation as in Section \ref{contspecasym}.

We consider the ad hoc approximation of the infinite beam impulse response $\bar u_{\infty,a}(\tau)$ as
\begin{equation}
	\bar u_{\infty,a}(\tau)
	=
	\bar P
	\sqrt{
		\sum_{i=1}^n
		\dfrac{a_i\tau^{i+1}}{1+b_i\tau^i}
	}.
	\label{eqapproxmain}
\end{equation}
This form is chosen to reproduce the known asymptotic structure of the Rayleigh beam impulse response: $\bar u_{\infty,a}=\mathcal{O}(\tau)$ for $\tau\ll1$ and $\bar u_{\infty,a}=\mathcal{O}(\sqrt{\tau})$ for $\tau\gg1$. The unknown coefficients are determined by expanding Eq.\ \eqref{eqapproxmain} in these two limits and matching coefficients with the corresponding asymptotic expansions of Section \ref{contspecasym}. The number of terms $n$ can be increased if desired. Here, we consider $n=2$ and $n=3$.

For $n=2$, we consider
\begin{equation}
	\bar u_{\infty,a}(\tau)
	=
	\bar P
	\sqrt{
		\dfrac{
			a_1\tau^2+a_2\tau^3
		}{
			1+b_1\tau+b_2\tau^2
	}}.
	\label{eqrational2}
\end{equation}
Matching the first two terms of the short-time expansion in Eq.\ \eqref{eqshortcont} and the first two terms of the large-time expansion in Eq.\ \eqref{eqlongcont} gives
\begin{equation}
	\begin{aligned}
		a_1&=\frac{1}{4}, &
		a_2&=\frac{2-\pi}{-8+\pi},\
		b_1&=-\frac{3\pi}{-8+\pi}, &
		b_2&=\frac{4\pi-2\pi^2}{-8+\pi}.
	\end{aligned}
	\label{eqrational2a}
\end{equation}
These coefficients are obtained using Mathematica. The resulting approximation captures both the initial linear growth and the long-time $\sqrt{\tau}$ behaviour while retaining a simple algebraic form. Increasing $n$ improves the approximation. 

For $n=3$, we consider
\begin{equation}
	\bar u_{\infty,a}(\tau)
	=
	\bar P
	\sqrt{
		\dfrac{
			a_1\tau^2+a_2\tau^3+a_3\tau^4
		}{
			1+b_1\tau+b_2\tau^2+b_3\tau^3
	}}.
	\label{eqrational}
\end{equation}
Now we can match the first three terms each of the short- and large-time asymptotic expansions. This gives
\begin{equation}
	\begin{aligned}
		a_1&=\frac{1}{4}, &
		a_2&=\frac{256-157\pi+16\pi^2}
		{-1024+208\pi+8\pi^2},&\
		a_3&=\frac{-24+31\pi-8\pi^2}
		{-128+26\pi+\pi^2},\\
		b_1&=\frac{18\pi^2-105\pi}
		{-256+52\pi+2\pi^2}, &
		b_2&=\frac{208\pi-95\pi^2}
		{-512+104\pi+4\pi^2},&\
		b_3&=\frac{-48\pi+62\pi^2-16\pi^3}
		{-128+26\pi+\pi^2}.
	\end{aligned}
	\label{eqrationala}
\end{equation}
The coefficients are again obtained using Mathematica. Numerical assessment of the resulting approximation is presented in Section \ref{comparison}.


\subsection{Numerical comparisons}
\label{comparison}

The direct numerical evaluation of the spectral integral in Section~\ref{approxinf} and the ad hoc approximations in Section~\ref{sqrtapprox} are compared with the exact analytical impulse response in Eq.~\eqref{eqexactinf} for the infinite Rayleigh beam at the point of application, with $\bar P=1$. For simplicity, the exact solution is denoted by $\bar u_{\rm an}(0,\tau)$, while the direct numerical solution and the two ad hoc approximations ($n=2$ and $n=3$) are collectively denoted by $\bar u\,(0,\tau)$.

\begin{figure}[htbp]
	\centering
	\includegraphics[width=\textwidth]{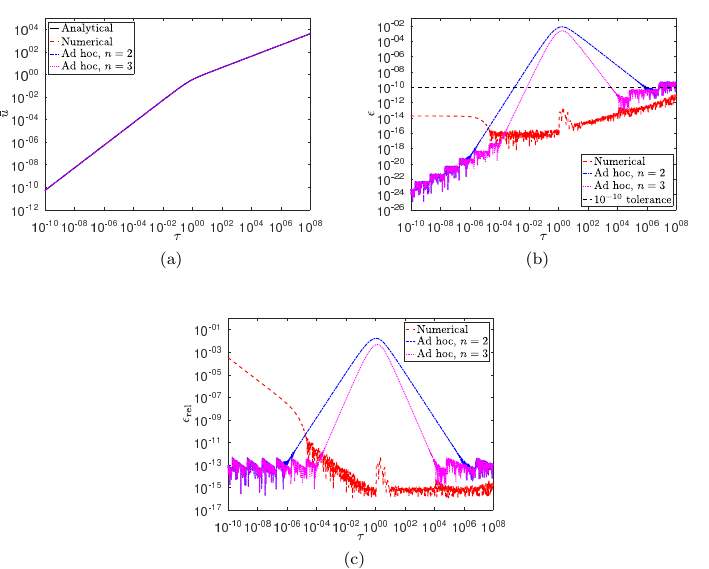}
	\caption{(a) Analytical impulse response together with the numerical integral, and the ad hoc approximations for $n=2,3$. (b) Pointwise absolute error $\epsilon$, for the direct numerical and approximate solutions. (c) Pointwise relative error $\epsilon_{\rm rel}$. $\bar P=1$.}
	\label{fig:comparison_methods}
\end{figure}

The numerical solution is obtained by evaluating the spectral integral
in Eq.\ \eqref{eqdef} as
described in Section \ref{approxinf}. The calculation covers the time
range $10^{-10}\leq\tau\leq10^8$, using $2000$ logarithmically spaced points in $\tau$, with a chosen absolute tolerance
$\varepsilon=10^{-10}$. The two ad hoc approximations are evaluated directly from Eqs.\ \eqref{eqrational2}-\eqref{eqrational2a} and
\eqref{eqrational}-\eqref{eqrationala}.

Figure \ref{fig:comparison_methods}(a) shows the impulse response from
the analytical expression, the direct numerical evaluation, and the ad hoc approximations. The numerical solution and the ad hoc approximations are visually indistinguishable from the analytical response over the
chosen range of $\tau$.

To quantify the errors in the numerical integral and the ad hoc approximations, we define the pointwise error
\begin{equation}
	\epsilon(\tau)
	=
	\left|\bar u(0,\tau)-\bar u_{\rm an}(0,\tau)\right|
	\label{eqabsoluteerror}
\end{equation}
and the pointwise relative error
\begin{equation}
	\epsilon_{\rm rel}(\tau)
	=
	\left|
	\frac{\bar u(0,\tau)-\bar u_{\rm an}(0,\tau)}
	{\bar u_{\rm an}(0,\tau)}
	\right|.
	\label{eqrelativeerror}
\end{equation}
These are plotted in Figs.\ \ref{fig:comparison_methods}(b) and (c)
for the numerically evaluated integral and both the ad hoc approximations, respectively.

Figure \ref{fig:comparison_methods}(b) shows that the numerical
integral error stays below the chosen tolerance,
$\epsilon(\tau)\leq \varepsilon=10^{-10}$, across the full range $10^{-10}\leq\tau\leq10^8$.

The ad hoc approximations are, as expected, less accurate than the direct numerical evaluation, especially for $\tau = \mathcal{O}(1)$. Even so, they track the analytical solution closely. The maximum relative errors of the ad hoc approximations are found to be $1.64\%$ and $0.46\%$ for $n=2$ and $3$, respectively, with $n=3$ giving better agreement.

\section{Finite beam corrections to the infinite beam response}
\label{infval}

The infinite beam response describes the local response in the absence of the boundaries, whereas the finite beam response includes their effects arising from the two simply supported ends. The difference between the two responses has three distinct features. First, the finite-length correction is nonzero for every $\tau>0$, although it becomes exponentially small as the beam length increases. This follows from the short-time expansions in Eqs.~\eqref{eq28_full} and \eqref{eq29_full}, where the finite beam terms contain hyperbolic functions of $\bar L$ that decay rapidly with increasing beam length but remain technically nonzero for any finite $\bar L$. Second, because the Rayleigh beam is dispersive, wave packets travel with different group velocities. The fastest wave packets, corresponding to the maximum of the group velocity, are the first to return from the supports and produce a gradual, dispersive deviation of the finite beam response from the infinite beam response. Third, the highest frequency components propagate at a limiting velocity and form a wavefront. The return of this wavefront from the supports produces a discontinuous change in the slope of the velocity response. These three effects are examined below.

\subsection{Finite-length correction}
\label{finlen}

\begin{figure}[htbp]
	\centering
	\includegraphics[width=0.7\textwidth]{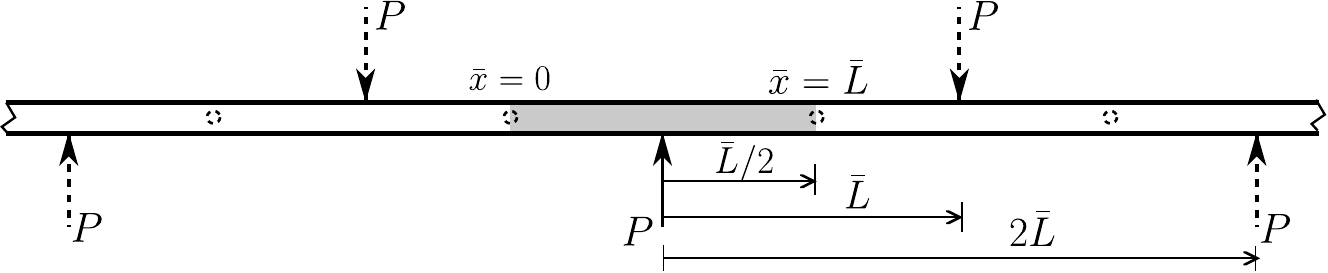}
	\caption{Image construction for a midpoint impulse in a simply supported beam. The locations denoted by dashed circles on the extended beam do not move, and act as if invisible hinges are present to satisfy the boundary conditions. Physical impulse is shown with solid arrow, image impulses are shown with dashed arrows. If all the impulses act simultaneously on the infinite beam, the response of the shaded portion is the same as that of a simply supported finite length beam.}
	\label{fig:imagemethod}
\end{figure}

For a finite beam with simply supported ends, $\bar u=\bar u_{\bar x\bar x}=0$ at each hinge. Both conditions can be enforced by reflecting the physical impulse about a hinge with a change of sign, since the combined response of the physical impulse and its image has zero displacement and zero moment at the hinge. The finite beam can therefore be represented by extending the beam beyond the hinges and introducing images so that these boundary conditions are satisfied at the hinge locations. For an impulse at the midpoint $\bar L/2$, enforcing the conditions at both hinges requires successive reflections about $\bar x=0$ and $\bar x=\bar L$. The resulting image sources occur at separations $j\bar L$ from the physical source, with $j=\pm1,\pm2,\ldots$, and alternate in sign, as shown in Fig.~\ref{fig:imagemethod}.

The infinite beam response $\bar u_\infty(\bar x,\tau)$ at a point $\bar x$ away from the struck point is given by Eq.~\eqref{eqinfx}, and can be used directly to evaluate the contribution of each image at the struck point. The physical impulse is located at the observation point, so its contribution is $\bar u_\infty(0,\tau)$. An image located at a distance $j\bar L$ from the struck point produces there the response $\bar u_\infty(j\bar L,\tau)$. By linearity, the finite beam response at the midpoint is obtained by superposing the response of the physical impulse and those of all image impulses. The alternating signs generated by the reflections give a factor $(-1)^j$ for the images at separation $j\bar L$. Moreover, $\bar u_\infty$ is even in $\bar x$, so the images at $+j\bar L$ and $-j\bar L$ make equal contributions. Hence,
\begin{equation}
	\bar u_F\!\left(\frac{\bar L}{2},\tau\right)
	=
	\bar u_\infty(0,\tau)
	+
	2\sum_{j=1}^{\infty}
	(-1)^j\bar u_\infty(j\bar L,\tau).
\end{equation}
Note, we are denoting the finite beam response by $\bar u_F$ here to distinguish from the infinite beam response $\bar u_\infty$. The finite-length correction $R_u$ is therefore the contribution of the image impulses,
\begin{equation}
	R_u(\tau,\bar L)
	=
	\bar u_F\!\left(\frac{\bar L}{2},\tau\right)
	-\bar u_\infty(0,\tau)
	=
	2\sum_{j=1}^{\infty}
	(-1)^j\bar u_\infty(j\bar L,\tau)
	=
	\frac{2\bar P}{\pi}
	\sum_{j=1}^{\infty}(-1)^j
	\int_0^\infty
	\frac{
		\sin\!\left(
		\dfrac{\kappa^2}{\sqrt{1+\kappa^2}}\,\tau
		\right)
		\cos(j\bar L\kappa)
	}{
		\kappa^2\sqrt{1+\kappa^2}
	}\,{\rm d}\kappa,
	\label{eqimg7a}
\end{equation}
where the last form follows by substituting Eq.~\eqref{eqinfx} at the image separation $\bar x=j\bar L$. The same image series of Eq.\ \eqref{eqimg7a} can also be derived formally from the modal expansion through Poisson summation (see Appendix~\ref{finlender}).

Equation~\eqref{eqimg7a} shows that the correction is present even at arbitrarily short times. A short-time asymptotic expansion of the right-hand side of Eq.~\eqref{eqimg7a} can be developed with increasing care at successive orders. The first term is obtained directly from the leading term of the sine expansion, since the resulting integral is sufficiently well behaved. Writing $R_u=\bar P\left(C_1\tau+C_3\tau^3+\cdots\right)$, the coefficient of the linear term is
\begin{equation}
	C_1(\bar L)
	=
	\frac{2}{\pi}\sum_{j=1}^{\infty}(-1)^j
	\int_0^\infty\frac{\cos(j\bar L\kappa)}{1+\kappa^2}\,{\rm d}\kappa
	=
	\sum_{j=1}^{\infty}(-1)^j e^{-j\bar L}
	=
	-\frac{1}{e^{\bar L}+1}
	=
	\frac{1}{2}\left[\tanh\left(\frac{\bar L}{2}\right)-1\right],
	\label{eqshortC1}
\end{equation}
where we used $\displaystyle \int_0^\infty\frac{\cos(a\kappa)}{1+\kappa^2}\,{\rm d}\kappa
=\frac{\pi}{2}e^{-a}$, for $a>0$. The next term is more delicate because the corresponding integral is not absolutely convergent. If one integral is interpreted in the regularized sense, the same procedure gives
\begin{equation}
	C_3(\bar L)
	=
	\frac{1}{3\pi}\sum_{j=1}^{\infty}(-1)^{j+1}
	\int_0^\infty
	\frac{\kappa^4}{(1+\kappa^2)^2}\cos(j\bar L\kappa)\,{\rm d}\kappa
	=
	\frac{\bar L}{48}\operatorname{sech}^2\left(\frac{\bar L}{2}\right)
	-
	\frac{1}{8}\left[1-\tanh\left(\frac{\bar L}{2}\right)\right].
	\label{eqflC3}
\end{equation}
Here we have used $\displaystyle \frac{\kappa^4}{(1+\kappa^2)^2}
=
1-\frac{2}{1+\kappa^2}
+\frac{1}{(1+\kappa^2)^2},
$ together with $\displaystyle \int_0^\infty
\frac{\cos(a\kappa)}{(1+\kappa^2)^2}\,{\rm d}\kappa
=
\frac{\pi}{4}(1+a)e^{-a}$, for $a>0$, and the regularized integral $\displaystyle \int_0^\infty\cos(a\kappa)\,{\rm d}\kappa=0$.

The coefficients of $\tau$ and $\tau^3$, obtained from the short-time asymptotic expansion of Eq.~\eqref{eqimg7a}, when added to the corresponding terms in Eq.~\eqref{eq29_full}, correctly reproduce upto $\mathcal{O}(\tau^3)$ of the finite beam expansion in Eq.~\eqref{eq28_full}. Higher-order terms require a more careful asymptotic treatment and are not pursued here. This agreement confirms that the finite-length correction is consistent with the short-time asymptotic expansion of the finite beam response. The finite-length correction is thus present from arbitrarily small times, while the boundary-reflection effects, considered next, become appreciable some finite time interval later.


\subsection{Boundary-reflection effects}
\label{groupvel}

\begin{figure}[htbp]
	\centering
	\includegraphics[width=0.4\textwidth]{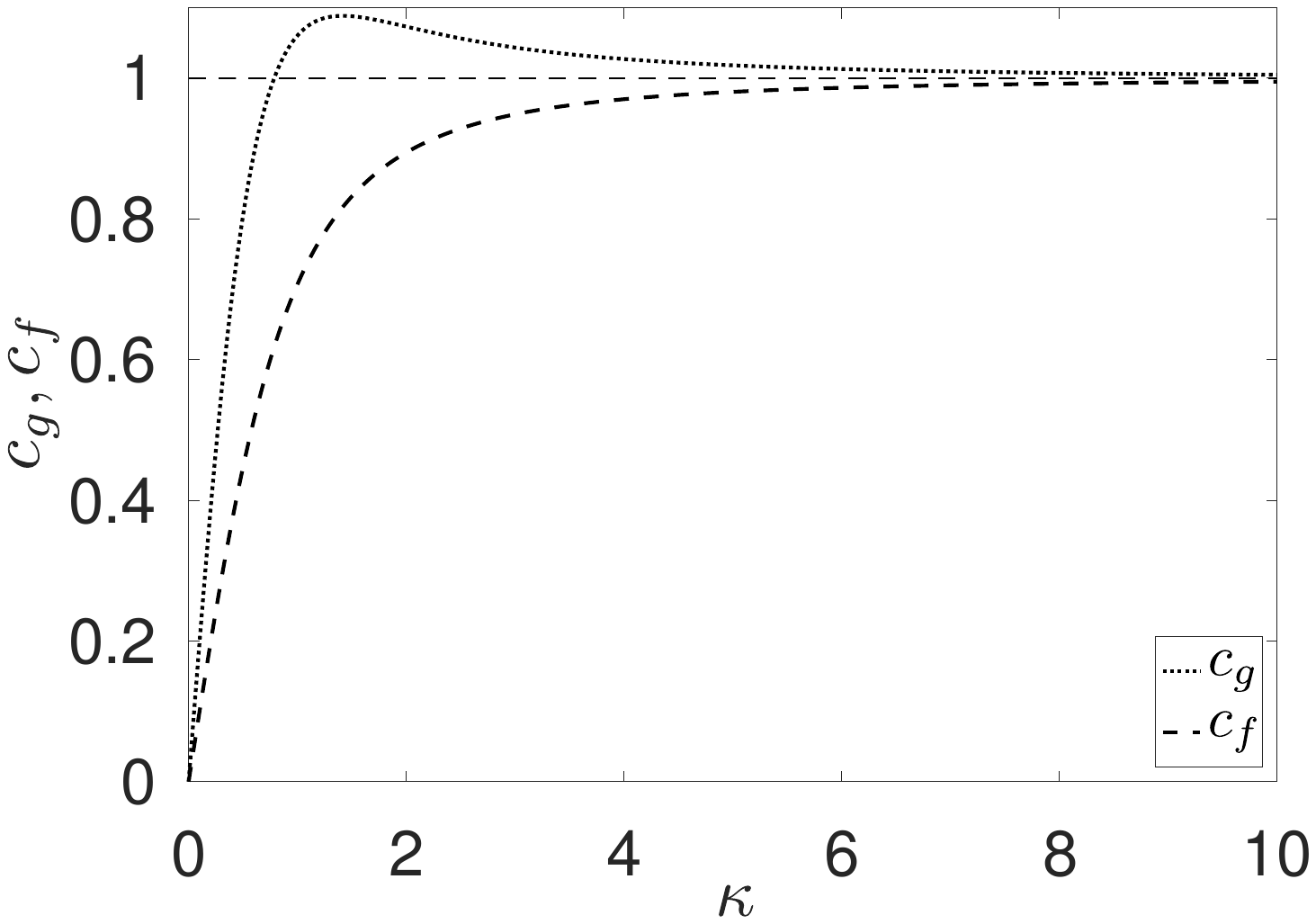}
	\caption{Group and phase velocities of the Rayleigh beam. The group velocity
		reaches a maximum at $\displaystyle\kappa=\sqrt{2}$ and subsequently approaches the
		limiting value $1$ from above, while the phase velocity approaches the
		same limit from below.}
	\label{fig:groupvel}
\end{figure}

The characteristic times associated with the boundary-induced response follow from the dispersion relation of the Rayleigh beam. A harmonic disturbance
$\displaystyle\bar u=\bar u_0\,e^{i(\kappa\bar x-\omega\tau)}$ in the homogeneous form of Eq.~\eqref{eq1nond} gives
\begin{equation}
	\omega^2
	=
	\frac{\kappa^4}{1+\kappa^2}.
	\label{eqvaldisp}
\end{equation}
The corresponding phase velocity is
\begin{equation}
	c_f
	=
	\frac{\omega}{\kappa}
	=
	\frac{\kappa}{\sqrt{1+\kappa^2}},
	\label{eqvalvelocitiesa}
\end{equation}
and the group velocity, which governs the
propagation of wave packets of nearby wavenumbers (see \cite{whitham2011linear}), is
\begin{equation}
	c_g
	=
	\frac{{\rm d}\omega}{{\rm d}\kappa}
	=
	\frac{\kappa(\kappa^2+2)}
	{(1+\kappa^2)^{3/2}}.
	\label{eqvalvelocities}
\end{equation}
The phase velocity increases monotonically with $\kappa$, whereas the group velocity has a maximum at
\begin{equation}
	\kappa_*=\sqrt{2},
	\qquad
	c_{g,\max}
	=
	\left.c_g\right\vert_{\kappa=\sqrt{2}}
	=
	\frac{4\sqrt6}{9}
	\approx 1.089.
	\label{eqvalcgmax}
\end{equation}
Both velocities approach unity at large wavenumber,
\begin{equation}
	\lim_{\kappa\rightarrow\infty}c_f
	=
	\lim_{\kappa\rightarrow\infty}c_g
	=
	1,
	\label{eqvalcf}
\end{equation}
with $c_f$ approaching the limit from below and $c_g$ from above, as shown in Fig.~\ref{fig:groupvel}.

The first detectable boundary-reflected disturbance comes from the nearest image
impulse, located a distance $\bar L$ from the observation point. For a
component with wavenumber $\kappa$, the group-velocity
travel time
\begin{equation}
	\tau_g(\kappa) = \frac{\bar L}{c_g(\kappa)} .
\end{equation}
This time is smallest where the group velocity is largest. Since $c_g$
reaches its maximum $c_{g,\max}$ at $\kappa_*=\sqrt{2}$,
\begin{equation}
	\tau_{\rm start}
	= \frac{\bar L}{c_{g,\max}}
	= \frac{9}{4\sqrt6}\,\bar L
	\approx 0.9186\,\bar L 
	\label{eqvaltstart}
\end{equation}
is the arrival time of the fastest wave packet according to the group velocity description. At $\kappa_*$, the group velocity is maximum, with $c_g'(\kappa_*)=\omega''(\kappa_*)=0$ and $c_g''(\kappa_*)=\omega'''(\kappa_*)=-2\sqrt6/27\neq0$. These are the local conditions associated with a dispersive wavefront, where the transition is described by
an Airy function \cite[\S13.6]{whitham2011linear},
\cite[\S1,3]{clarisse1996integrals}. Three results follow from this Airy description.
First, the transition is gradual. Although the leading-order
wavefront identifies $\tau_{\rm start}$ as a precise arrival time, the full
solution varies smoothly near $\tau_{\rm start}$
\cite[\S13.5]{whitham2011linear}. Second, the transition begins before $\tau_{\rm start}$. Ahead of the
front, the solution is exponentially decaying and non-oscillatory
\cite[\S13.5]{whitham2011linear}, while behind the front the Airy
description is oscillatory \cite[\S13.6]{whitham2011linear}. The nearest-image contribution can
therefore deviate slightly from the infinite beam response for
$\tau<\tau_{\rm start}$.
Third, if we set a small threshold for a detectable change occurring at some 
$\tau<\tau_{\rm start}$ then, as per theory,
\begin{equation}
	|\tau-\tau_{\rm start}|=\mathcal O(\bar L^{1/3}),
	\label{eqvaltransitionwidth}
\end{equation}
consistent with the width $\propto t^{1/3}$ of the Airy transition
\cite[\S13.6]{whitham2011linear}, and with the Airy argument
$\propto t^{2/3}$ times the distance from the front in \cite[Eq.~(3.13)]{clarisse1996integrals}.
Equation \eqref{eqvaltransitionwidth} will be verified numerically for our specific problem in the next section.

The dispersion relation also gives a second characteristic time. In the
high-wavenumber limit, both the group and phase velocities approach unity, so
that the fully developed wavefront, which contains infinitely many
wavenumber components, travels the distance $\bar L$ in time
\begin{equation}
	\tau_{\rm lim}
	=
	\bar L.
	\label{eqvaltref}
\end{equation}
Thus,
\begin{equation}
	\tau_{\rm start}
	=
	\frac{9}{4\sqrt6}\,\tau_{\rm lim}
	\approx0.9186\,\tau_{\rm lim}.
	\label{eqvaltstartref}
\end{equation}
Unlike the arrival at $\tau_{\rm start}$, which involves only a narrow band of
wavenumbers near $\kappa_*$, the arrival at $\tau_{\rm lim}$ includes an infinite range of
high-wavenumber components.  Technically, the contribution from progressively higher wavenumbers is concentrated
arbitrarily close to $\tau_{\rm lim}$ rather than being distributed over a
finite transition interval. The limiting high-wavenumber contribution
therefore produces a sharp change in slope of the velocity response at
$\tau_{\rm lim}$, in contrast to the finite-width transition produced by
the wavenumbers near $\kappa_*$ at $\tau_{\rm start}$.


\subsection{Final numerical observations}
\label{validation}

\begin{figure}[p]
	\centering
	\includegraphics[width=0.48\textwidth,height=0.28\textheight,keepaspectratio]{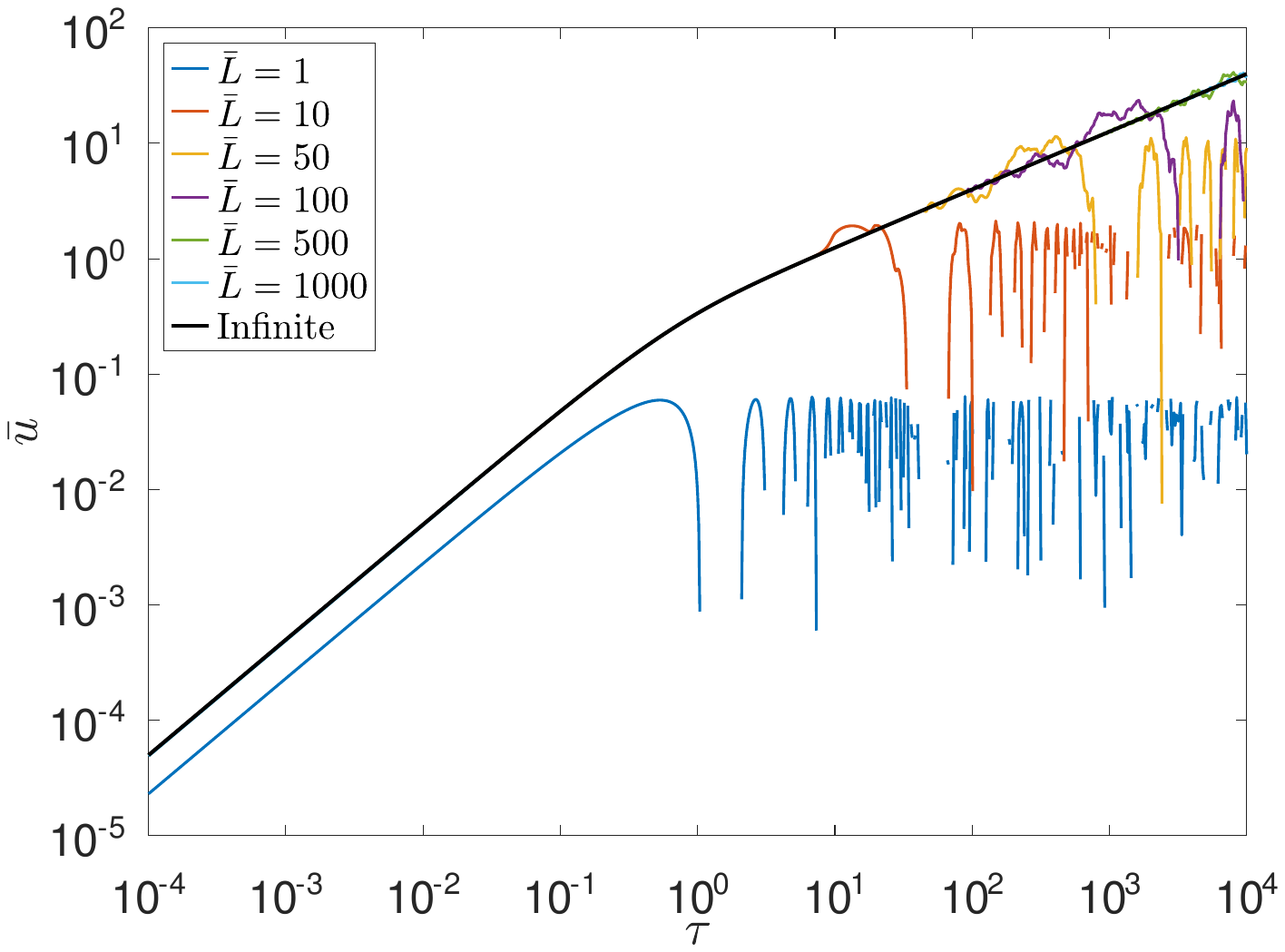}
	\caption{Finite beam response at the point of application for increasing
		nondimensional beam length $\bar L=1,10,50,100,500,1000$, compared with the
		infinite beam response; for $\bar P=1$.}
	\label{figlcomp}
	
	\vspace{1.5em}
	
	\includegraphics[width=0.85\textwidth,keepaspectratio]{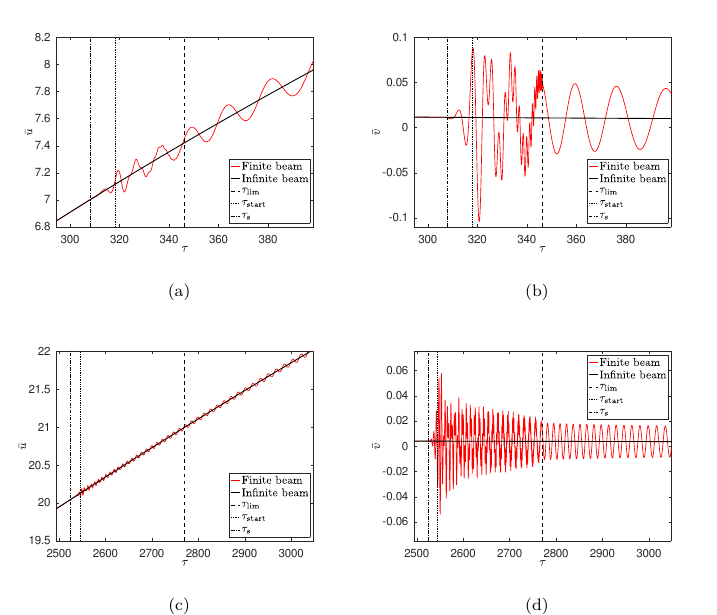}
	\caption{Finite and infinite beam responses for steel beams with 1 cm $\times$ 1 cm cross section. (a) $\bar L=346.4$, displacement response. (b) $\bar L=346.4$, velocity response. (c) $\bar L=2771.2$,  displacement response. (d) $\bar L=2771.2$,  velocity response. Time $\tau_{\rm start}$ corresponds to the arrival of a reflection at the maximum group velocity; $\tau_{\rm s} < \tau_{\rm start}$ corresponds to a detectable deviation in velocity; and $\tau_{\rm lim}$ corresponds to the simultaneous arrival of reflections at arbitrarily high-frequencies. The finite beam responses were computed using $10^5$ modes in all cases.}
	\label{figlcompsteel}
\end{figure}

The analysis above points to some characteristic features of the finite beam response and their dependence on the beam length. These features are demonstrated numerically below.

Figure~\ref{figlcomp} compares the finite beam response from Eq.~\eqref{eqmod8} with the infinite beam response from Eq.~\eqref{eqexactinf} for several values of $\bar L$. The responses become progressively closer to the infinite beam response as $\bar L$ increases. In particular, the difference in the power series itself (Eqs.\ \eqref{eq28_full} and \eqref{eq29_full}), which extends all the way to
$\tau \rightarrow 0^+$, is easily visible for $\bar L = 1$ but too small to discern visually for larger $\bar L$. Even for larger $\bar L$, however, the response eventually deviates for large enough $\tau$ (larger $\tau$ for larger $\bar L$). These observations are consistent with the foregoing discussion.
We note that $\bar L = 1$ corresponds to an extremely short bar wherein the beam modelling assumptions may be inappropriate, but the behaviour of solutions to the underlying PDE remains of theoretical interest.

Next, the boundary-reflection effects are illustrated using parameters corresponding to two steel beams in
Fig.~\ref{figlcompsteel}. These beams have $E=210$ GPa and
$\rho=7850$ kg/m$^3$, differing physical lengths $L=1$ m and
$8$ m, and identical 1 cm square cross sections with $A=10^{-4}$ m$^2$ and
$I=8.33\times10^{-10}$ m$^4$. These parameters give
$\bar L=L\sqrt{A/I}=346.4$ and $2771.2$, respectively\footnote{%
	Thus, a 1 cm sized cubical object would have $\bar L = 3.464$, so $\bar L = 1$ is clearly unphysical.}.
The dimensionless
time is $\tau=t\omega_0$, where
$\omega_0=\sqrt{EA/\rho I}=1.79\times10^{6}\,\mathrm{s^{-1}}$.

In Fig.~\ref{figlcompsteel}(a), for $\bar{L}=346.4$, the finite and infinite beam displacement responses are
initially almost indistinguishable, although a small gradual departure becomes
visible before $\tau_{\mathrm{start}}$. The departure becomes more pronounced
with increasing time. In Fig.~\ref{figlcompsteel}(b) the velocity response resolves this same deviation more
clearly, and a sharper change is clearly visible near
$\tau_{\mathrm{lim}}=\bar{L}$ after some high-frequency oscillations. The sudden cessation of high frequency oscillations after $\tau_{\mathrm{lim}}$ corresponds to the high-frequency reflected wave train having passed the struck point.

To examine the response near $\tau_{\rm start}$, we arbitrarily choose  our  detection threshold using velocities as
\begin{equation}
	\frac{\left|\bar{v}-\bar{v}_{\infty}\right|}{\left|\bar{v}_{\infty}\right|}\geq 0.05 .
	\label{eq:cutoffvres}
\end{equation}
For $\bar{L}=346.4$, $\tau_{\mathrm{start}}=318.19$, while the first instant when the threshold is crossed is $\tau_{\mathrm{s}}=308.14$, giving
$\tau_{\mathrm{start}}-\tau_{\mathrm{s}}=10.05$. For the longer beam with
$\bar{L}=2771.2$, $\tau_{\mathrm{start}}=2545.51$ and
$\tau_{\mathrm{s}}=2524.96$, giving
$\tau_{\mathrm{start}}-\tau_{\mathrm{s}}=20.55$. Thus, increasing $\bar{L}$ by
a factor of eight increases the separation between the detectable onset and
$\tau_{\mathrm{start}}$ by a factor of 
$20.55/10.05 = 2.04$, consistent with the factor $8^{1/3}=2$ predicted by the
transition-width scaling $\mathcal{O}(\bar{L}^{1/3})$ of
Eq.~\eqref{eqvaltransitionwidth}.

We close this section with a possible practical application of the foregoing analysis. Consider an impact of a compact body like a Hertzian sphere on a long, slender beam, with the point of impact sufficiently far from the boundaries. If the contact duration is found, a posteriori, to be smaller than the detectable onset of boundary-reflection effects, then the finite beam may be replaced with the corresponding infinite beam model with high accuracy. The infinite beam formulation provides a simpler alternative for finite beam impact problems, and the results obtained above give us quantitative criteria for deciding whether the simplification is valid.


\section{Concluding remarks}
\label{conc}

The Rayleigh beam studied here extends the classical Euler-Bernoulli model by including rotary inertia. This makes the model more realistic, but also makes the problem analytically more involved and leads to behaviour that is qualitatively different from the Euler-Bernoulli case. In particular, the response exhibits distinct short- and long-time behaviours; and the wave propagation, though dispersive, occurs at bounded speeds.

While the investigations in this paper have relied on several tools of classical applied mathematics, including modal expansion, Fourier and Laplace transforms, asymptotic analyses, numerical contour integration with pre-specified error bounds, and some ad hoc approximations as well, the full problem has yielded a satisfactory level of closure. The problem and its solution therefore offer some academic value for researchers and students of dynamics including vibrations and waves.

It was found that the difference between finite and infinite beam responses can be understood largely in terms of three distinct effects associated with reflections from boundaries. Of these three effects, the first one extends to arbitrarily small times but decays exponentially with beam length and is negligible for physically realistic beam geometries. The other two effects arise from (i) a local maximum in group velocity with an associated small precursor that is consistent with more general theoretical results in the literature, and (ii) a sharp wavefront corresponding to the near-simultaneous arrival of infinitely many high frequency components. In particular, the full time series computed from modal expansion satisfactorily shows the arrival and passage of such high frequency components (Fig.~\ref{figlcompsteel}(b)).

From a practical point of view, this paper can lead to a study of impact between a Hertzian sphere and a long Rayleigh beam, which we hope to pursue in future work as a follow-up to the study of \cite{BHATTACHARJEE2018147}.


\appendix

\section*{Appendix}
\section{Derivation of the finite beam short-time asymptotics} \label{appn_asymcal}
We present here the detailed derivation of the asymptotic expansion of Eq.\ \eqref{eqmod8}. The procedure follows the general asymptotic framework of Goswami \textit{et al.}\ \cite{goswami2023short}, adapted to the present problem. The derivation makes repeated use of the Euler-Maclaurin summation formula,

\begin{equation}
	\sum_{n=N+1}^{\infty} g(n) = \int_N^\infty g(y) \mathrm{d}y - \dfrac{g(N)}{2}-\dfrac{g'(N)}{12}+\dfrac{g'''(N)}{720}-\dfrac{g^{(5)}(N)}{30240}+ \dfrac{g^{(7)}(N)}{1209600}\cdots ,
	\label{appneq12}
\end{equation}
where the prime denotes differentiation with respect to $y$. We have used Mathematica for symbolic computations.

Let us consider the infinite sum in Eq.\ \eqref{eqmod8} again as
\begin{equation}
	\bar u\left(\dfrac{\bar L}{2},\tau\right)=\dfrac{2 \bar P}
	{\bar L}\sum_{n=1}^{\infty}  \frac{1}{(2n-1)^2 q^2 \sqrt{1+(2n-1)^2 q^2}} \sin \left(\dfrac{(2n-1)^2 q^2}{\sqrt{1+(2n-1)^2 q^2}} \, \tau\right), \quad \mathrm{for} \,\, \, \, 0<\tau\ll 1.
	\label{eq9}
\end{equation}
Equation \eqref{eq9} can be recast in the form of
\begin{equation}
	S(\tau)=\dfrac{2 \bar P}
	{\bar L}\sum_{n=1}^{N} f(n,\tau) + \dfrac{2 \bar P}
	{\bar L}\sum_{n=N+1}^{\infty} f(n,\tau),
	\label{eq10}
\end{equation}
where $f(n,\tau)=\dfrac{1}{(2n-1)^2 q^2 \sqrt{1+(2n-1)^2 q^2}} \sin \left(\dfrac{(2n-1)^2 q^2}{\sqrt{1+(2n-1)^2 q^2}} \, \tau\right)$, and $N$ is a large number on the order of $N \sim \dfrac{1}{\sqrt{\tau}}$. Assuming that the asymptotic dependence of relevant quantities is smooth in $\tau$, we can use discrete values of $\tau$ (our assumption is amply borne out by numerics), and our analysis will be simplified if we take $N = \left(\dfrac{1}{\sqrt{\tau}}+\dfrac{1}{2}\right)$. Equation \eqref{eq10} can further be rewritten in the form of
\begin{equation}
	S(\tau)=\dfrac{2 \bar P}
	{\bar L} S_1 + \dfrac{2 \bar P}
	{\bar L}S_2,
	\label{eq11}
\end{equation}
where $\displaystyle S_1=\sum_{n=1}^{N} f(n,\tau)$ and $\displaystyle S_2=\sum_{n=N+1}^{\infty} f(n,\tau)$. We will evaluate $S_1$ and $S_2$ separately and combine them to find $S(\tau)$.

\subsection{Expression for $S_1$}
We begin by evaluating
\begin{equation}
	S_1=\sum_{n=1}^{N} \frac{1}{(2n-1)^2 q^2 \sqrt{1+(2n-1)^2 q^2}} \sin \left(\dfrac{(2n-1)^2 q^2}{\sqrt{1+(2n-1)^2 q^2}} \, \tau\right).
	\label{eq13_full}
\end{equation}
For $n<N$ and $\tau\ll 1$, Eq.\ \eqref{eq13_full} can be expanded in a Taylor series as
\begin{equation}
	\begin{split}
		S_1&=\sum_{n=1}^{N} \dfrac{\tau}{(1+(2n-1)^2q^2)} - \sum_{n=1}^{N} \dfrac{(2n-1)^4q^4\tau^3}{6(1+(2n-1)^2q^2)^2} + \sum_{n=1}^{N} \dfrac{(2n-1)^8q^8\tau^5}{120(1+(2n-1)^2q^2)^3} \\
		&- \sum_{n=1}^{N} \dfrac{(2 n-1)^{12} q^{12} \tau^7}{5040 \left((2 n-1)^2 q^2+1\right)^4} + \sum_{n=1}^{N} \dfrac{(2 n-1)^{16} q^{16} \tau ^9}{362880 \left((2 n-1)^2 q^2+1\right)^5}+ \mathcal{O}\left(\tau^{7}\right),
	\end{split}
	\label{appneq14_full}
\end{equation}
where we have accounted for the largest $n$ values to write the error term.
Define $S_{11}$, $S_{12}$, $S_{13}$, and $S_{14}$ as the first 4 terms of the RHS of Eq.\ \eqref{appneq14_full}, respectively. Thus, dropping the smallest term now that we know the pattern,
\begin{equation}
	S_1 = S_{11}-S_{12}+S_{13}-S_{14} + \mathcal{O}\left(\tau^6\right).
	\label{appneq15_full}
\end{equation}
Now, $S_{11}$ can be written as
\begin{eqnarray}
	S_{11} &=&\sum_{n=1}^{N} \dfrac{\tau}{(1+(2n-1)^2q^2)} = \tau \sum_{n=1}^{\infty} \dfrac{1}{(1+(2n-1)^2q^2)}-\tau \sum_{n=N+1}^{\infty} \dfrac{1}{(1+(2n-1)^2q^2)} \nonumber \\
	&=& \dfrac{\pi \tanh{\left(\dfrac{\pi}{2q}\right)}}{4q}\tau-\tau\sum_{n=N+1}^{\infty} \dfrac{1}{(1+(2n-1)^2q^2)}.
	\label{appneq16_full}
\end{eqnarray}
Using the Euler-Maclaurin expansion (Eq.\ \eqref{appneq12}), we have
\begin{equation}
	\begin{split}
		S_{11}&= \dfrac{\pi \tanh{\left(\dfrac{\pi}{2q}\right)}}{4q}\tau-\Bigg(\int_N^\infty \dfrac{\mathrm{d}y}{1+(2y-1)^2q^2}-\left.\dfrac{1}{2}\dfrac{1}{(1+(2y-1)^2q^2)}\right\vert_N\\
		&-\left.\dfrac{1}{12}\dfrac{\partial}{\partial y}\dfrac{1}{(1+(2y-1)^2q^2)}\right\vert_N
		+\left.\dfrac{1}{720}\dfrac{\partial^3}{\partial y^3}\dfrac{1}{(1+(2y-1)^2q^2)}\right\vert_N
		-\left.\dfrac{1}{30240}\dfrac{\partial^5}{\partial y^5}\dfrac{1}{(1+(2y-1)^2q^2)}\right\vert_N \\
		&+ \left.\dfrac{1}{1209600}\dfrac{\partial^7}{\partial y^7}\dfrac{1}{(1+(2y-1)^2q^2)}\right\vert_N + \cdots \Bigg)\,\tau.
	\end{split}
	\label{appneq17_full}
\end{equation}
After evaluating derivatives and substituting $N=\left(\dfrac{1}{\sqrt{\tau}}+\dfrac{1}{2}\right)$, the series expansion for small $\tau$ gives
\begin{equation}
	\begin{aligned}
		S_{11}= &\dfrac{\pi \tanh{\left(\dfrac{\pi}{2q}\right)}}{4q}\tau - \dfrac{\tau^{3/2}}{4q^2}+\dfrac{\tau^2}{8q^2} +\left(\dfrac{1}{48 q^4}-\dfrac{1}{24 q^2}\right) \tau^{5/2} -\dfrac{\tau^3}{32q^4}+ \left(  -\frac{1}{320 q^6}+\frac{1}{48 q^4}+\frac{1}{120 q^2} \right)\tau^{7/2} \\ &+ \dfrac{\tau^4}{128q^6} +\left(\dfrac{1}{1792 q^8}-\dfrac{1}{128 q^6}-\dfrac{1}{96 q^4}-\dfrac{1}{168 q^2}\right) \tau^{9/2}-\dfrac{\tau^5}{512 q^8} + \mathcal{O}\left(\tau^{11/2}\right).
	\end{aligned}
	\label{appneq18_full}
\end{equation}

Similarly, 
\begin{equation}
	\begin{split}
		S_{12}&=\sum_{n=1}^{N} \dfrac{(2n-1)^4q^4\tau^3}{6(1+(2n-1)^2q^2)^2} = \tau^3 \left(\sum_{n=1}^{N} \dfrac{1}{6}+\sum_{n=1}^{N} \dfrac{1}{6(1+(2n-1)^2q^2)^2} - \sum_{n=1}^{N} \dfrac{1}{3(1+(2n-1)^2q^2)}\right)\\
		&=\dfrac{N\tau^3}{6}+\tau^3\sum_{n=1}^{\infty} \dfrac{1}{6(1+(2n-1)^2q^2)^2}-\tau^3\sum_{n=1}^{\infty} \dfrac{1}{3(1+(2n-1)^2q^2)}-\tau^3\sum_{n=N+1}^{\infty} \dfrac{1}{6(1+(2n-1)^2q^2)^2}\\
		&+\tau^3\sum_{n=N+1}^{\infty} \dfrac{1}{3(1+(2n-1)^2q^2)}
	\end{split} 
	\label{appneq19_full}
\end{equation}
Evaluating the sums using Eq.\ \eqref{appneq12}, we obtain
\begin{equation}
	\begin{aligned}
		S_{12}&=\dfrac{\tau^{5/2}}{6}+\tau ^3 \left(-\dfrac{\pi ^2 \text{sech}^2\left(\dfrac{\pi }{2 q}\right)}{96 q^2}-\dfrac{\pi  \tanh \left(\dfrac{\pi }{2 q}\right)}{12 q}+\dfrac{\pi  \sinh \left(\dfrac{\pi }{q}\right) \text{sech}^2\left(\dfrac{\pi }{2 q}\right)}{96 q}+\dfrac{1}{12}\right)+\dfrac{\tau^{7/2}}{12 q^2}-\dfrac{\tau^4}{24 q^2} \\ & +\left(\dfrac{1}{72 q^2}-\dfrac{1}{96 q^4}\right) \tau ^{9/2}+\dfrac{\tau^5}{64 q^4}+\mathcal{O}\left(\tau^{11/2}\right).
	\end{aligned}
	\label{appneq20_full}
\end{equation}

Next, we consider
\begin{equation}
	\begin{split}
		S_{13}&=\sum_{n=1}^{N} \dfrac{(2n-1)^8q^8\tau^5}{120(1+(2n-1)^2q^2)^3} \\
		&= -\dfrac{N \tau^5}{40} +\dfrac{N(4N^2-1)q^2 \tau^5}{360}+\tau^5 \sum_{n=1}^{\infty}  \left(\dfrac{1}{120(1+(2n-1)^2q^2)^3}-\dfrac{1}{30(1+(2n-1)^2q^2)^2}+\dfrac{1}{20(1+(2n-1)^2q^2)} \right)\\
		&-\tau^5\sum_{n=N+1}^{\infty}\left(\dfrac{1}{120(1+(2n-1)^2q^2)^3}-\dfrac{1}{30(1+(2n-1)^2q^2)^2}+\dfrac{1}{20(1+(2n-1)^2q^2)} \right)
	\end{split}
	\label{appneq21_full}
\end{equation}
Evaluating using Eq.\ \eqref{appneq12} and expanding for small $\tau$ gives
\begin{equation}
	\begin{aligned}
		S_{13}&= \dfrac{1}{90} q^2 \tau^{7/2}+\dfrac{q^2 \tau^4}{60}+\left(\dfrac{q^2}{180}-\dfrac{1}{40}\right) \tau^{9/2}+\tau ^5 \Bigg(-\dfrac{\pi ^3 \tanh \left(\dfrac{\pi }{2 q}\right) \text{sech}^2\left(\dfrac{\pi }{2 q}\right)}{7680 q^3}+\dfrac{13 \pi ^2 \text{sech}^2\left(\dfrac{\pi }{2 q}\right)}{7680 q^2} \\ & +\dfrac{17 \pi  \tanh \left(\dfrac{\pi }{2 q}\right)}{1280 q}-\dfrac{\pi  \sinh \left(\dfrac{\pi }{q}\right) \text{sech}^2\left(\dfrac{\pi }{2 q}\right)}{480 q}-\dfrac{1}{80}\Bigg)+\mathcal{O}\left(\tau^{11/2}\right).
	\end{aligned}
	\label{appneq22_full}
\end{equation}

Finally, 
\begin{equation}
	\begin{split}
		S_{14}&=\sum_{n=1}^{N} \dfrac{(2 n-1)^{12} q^{12} \tau^7}{5040 \left((2 n-1)^2 q^2+1\right)^4} \\
		&= \dfrac{N \tau^7}{504} - \dfrac{(4 N^3 q^2-N q^2) \tau^7}{3780} + \dfrac{(48 N^5 q^4-40 N^3 q^4+7 N q^4) \tau^7}{75600} + \tau^7 \sum_{n=1}^{\infty} \Bigg \{-\dfrac{1}{252 \left(4 n^2 q^2-4 n q^2+q^2+1\right)} \\
		&+\dfrac{1}{336 \left(4 n^2 q^2-4 n q^2+q^2+1\right)^2} -\dfrac{1}{840 \left(4 n^2 q^2-4 n q^2+q^2+1\right)^3} +\dfrac{1}{5040 \left(4 n^2 q^2-4 n q^2+q^2+1\right)^4} \Bigg \} \\
		&- \tau^7 \sum_{n=N+1}^{\infty} \Bigg \{-\dfrac{1}{252 \left(4 n^2 q^2-4 n q^2+q^2+1\right)} + \dfrac{1}{336 \left(4 n^2 q^2-4 n q^2+q^2+1\right)^2} \\
		&-\dfrac{1}{840 \left(4 n^2 q^2-4 n q^2+q^2+1\right)^3} +\dfrac{1}{5040 \left(4 n^2 q^2-4 n q^2+q^2+1\right)^4} \Bigg \}.
	\end{split}
	\label{appneq221_full}
\end{equation}
Following similar methods as above, we obtain
\begin{equation}
	S_{14}=\dfrac{q^4 \tau ^{9/2}}{1575}+\dfrac{q^4 \tau ^5}{630}+\mathcal{O}\left(\tau^{11/2}\right).
	\label{eq222_full}
\end{equation}

Combining all contributions,
\begin{equation}
	\begin{split}
		S_1 &\sim S_{11}-S_{12}+S_{13}-S_{14}  \\
		&= \dfrac{\pi \tanh \left(\dfrac{\pi}{2q}\right) \tau}{4q} -\dfrac{\tau^{3/2}}{4q^2} + \dfrac{\tau^2}{8 q^2} +\left(\dfrac{1}{48 q^4}-\dfrac{1}{24 q^2}-\dfrac{1}{6}\right) \tau ^{5/2} \\
		&+\left(-\dfrac{1}{32 q^4}+\dfrac{\pi ^2 \text{sech}^2\left(\dfrac{\pi }{2 q}\right)}{96 q^2}+\dfrac{\pi  \tanh \left(\dfrac{\pi }{2 q}\right)}{12 q}-\dfrac{\pi  \sinh \left(\dfrac{\pi }{q}\right) \text{sech}^2\left(\dfrac{\pi }{2 q}\right)}{96 q}-\dfrac{1}{12}\right)\tau ^3 \\
		&+\left(-\dfrac{1}{320 q^6}+\dfrac{1}{48 q^4}+\dfrac{q^2}{90}-\dfrac{3}{40 q^2}\right) \tau ^{7/2} +\left(\dfrac{1}{128 q^6}+\dfrac{q^2}{60}+\dfrac{1}{24 q^2}\right) \tau ^4 +\Bigg(\dfrac{1}{1792 q^8}-\dfrac{1}{128 q^6}-\dfrac{q^4}{1575}+\dfrac{q^2}{180} \\ & -\dfrac{5}{252 q^2}-\dfrac{1}{40}\Bigg) \tau ^{9/2}+ \Bigg(-\dfrac{1}{512 q^8}-\dfrac{q^4}{630}-\dfrac{1}{64 q^4}-\dfrac{\pi ^3 \tanh \left(\dfrac{\pi }{2 q}\right) \text{sech}^2\left(\dfrac{\pi }{2 q}\right)}{7680 q^3}+\dfrac{13 \pi ^2 \text{sech}^2\left(\dfrac{\pi }{2 q}\right)}{7680 q^2} \\ & +\dfrac{17 \pi  \tanh \left(\dfrac{\pi }{2 q}\right)}{1280 q}-\dfrac{\pi  \sinh \left(\dfrac{\pi }{q}\right) \text{sech}^2\left(\dfrac{\pi }{2 q}\right)}{480 q}-\dfrac{1}{80}\Bigg) \tau ^5 + \mathcal{O}\left(\tau^{11/2}\right).
	\end{split}
	\label{appneq23_full}
\end{equation}

\subsection{Expression for $S_2$}
Next, we consider 
\begin{equation}
	S_2=\sum_{n=N+1}^{\infty} \frac{1}{(2n-1)^2 q^2 \sqrt{1+(2n-1)^2 q^2}} \sin \left(\dfrac{(2n-1)^2 q^2}{\sqrt{1+(2n-1)^2 q^2}} \, \tau\right).
	\label{appneq24_full}
\end{equation}
Let $2n-1=m$, which is also large. The summand of Eq.\ \eqref{appneq24_full} can be recast as
\begin{equation}
	\dfrac{1}{m^2 q^2 \sqrt{1+m^2 q^2}} \sin \left(\dfrac{m q}{\sqrt{1+\dfrac{1}{m^2 q^2}}} \, \tau\right).
	\label{appneq24_1_full}
\end{equation}
For large $m$, the argument is expanded as
\begin{equation}
	\dfrac{m q \tau}{\sqrt{1+\dfrac{1}{m^2 q^2}}} = m q \tau -\dfrac{\tau}{2 m q} +\dfrac{3 \tau}{8 m^3 q^3} -\dfrac{5 \tau}{16 m^5 q^5} +\dfrac{35 \tau}{128 m^7 q^7} -\dfrac{63 \tau}{256 m^9 q^9} +\dfrac{231 \tau }{1024 m^{11} q^{11}} +\mathcal{O}\left(m^{-12} \tau \right).
	\label{appneq24_2_full}
\end{equation}

We write 
\begin{equation}
	\theta=m q \tau, \qquad \mathrm{and} \qquad \delta=-\dfrac{\tau}{2 m q} +\dfrac{3 \tau}{8 m^3 q^3} -\dfrac{5 \tau}{16 m^5 q^5} +\dfrac{35 \tau}{128 m^7 q^7} -\dfrac{63 \tau}{256 m^9 q^9} +\dfrac{231 \tau }{1024 m^{11} q^{11}} +\mathcal{O}\left(m^{-12} \tau \right).
	\label{appneq24_2_1_full}
\end{equation}
Note that $|\delta| \ll \theta \ll 1$. An expansion for small $\delta$ gives
\begin{equation}
	\begin{aligned}
		\sin (\theta + \delta) = & \sin (\theta )+\delta  \cos (\theta )-\frac{1}{2} \delta ^2 \sin (\theta )-\frac{1}{6} \delta ^3 \cos (\theta )+\frac{1}{24} \delta ^4 \sin (\theta )+\frac{1}{120} \delta ^5 \cos (\theta ) \\ &-\frac{1}{720} \delta ^6 \sin (\theta )-\frac{\delta ^7 \cos (\theta )}{5040}+\mathcal{O}\left(\delta ^8\right).
	\end{aligned}
	\label{appneq24_3_full}
\end{equation}
Using Eqs.\ \eqref{appneq24_2_full}, \eqref{appneq24_2_1_full} and \eqref{appneq24_3_full}, the sine term of \eqref{appneq24_1_full} can be expressed as
\begin{equation}
	\begin{aligned}
		\sin \left(\dfrac{m q \tau}{\sqrt{1+\dfrac{1}{m^2 q^2}}} \, \right) & =  \sin (m q \tau)-\dfrac{\tau \cos (m q \tau)}{2 m q} -\dfrac{\tau^2 \sin (m q \tau)}{8 m^2 q^2}+\dfrac{\dfrac{\tau^3 \cos (m q \tau)}{48 q^3}+\dfrac{3 \tau \cos (m q \tau)}{8 q^3}}{m^3} \\
		& +\dfrac{\dfrac{\tau^4 \sin (m q \tau)}{384 q^4}+\dfrac{3 \tau^2 \sin (m q \tau)}{16 q^4}}{m^4} +\dfrac{-\dfrac{\tau^5 \cos (m q \tau)}{3840 q^5}-\dfrac{3 \tau^3 \cos (m q \tau)}{64 q^5}-\dfrac{5 \tau \cos (m q \tau)}{16 q^5}}{m^5} \\ & +\dfrac{-\dfrac{\tau^6 \sin (m q \tau)}{46080 q^6}-\dfrac{\tau^4 \sin (m q \tau)}{128 q^6}-\dfrac{29 \tau^2 \sin (m q \tau)}{128 q^6}}{m^6} \\ & +\dfrac{\dfrac{\tau ^7 \cos (m q \tau )}{645120 q^7}+\dfrac{\tau^5 \cos (m q \tau)}{1024 q^7}+\dfrac{19 \tau^3 \cos (m q \tau)}{256 q^7}+\dfrac{35 \tau \cos (m q \tau)}{128 q^7}}{m^7} \\ & 
		+\dfrac{\dfrac{\tau^6 \sin (m q \tau)}{10240 q^8}+\dfrac{47 \tau^4 \sin (m q \tau)}{3072 q^8}+\dfrac{65 \tau^2 \sin (m q \tau)}{256 q^8}}{m^8} \\ &
		+\dfrac{-\dfrac{\tau ^7 \cos (m q \tau )}{122880 q^9}-\dfrac{7 \tau^5 \cos (m q \tau)}{3072 q^9}-\dfrac{13 \tau^3 \cos (m q \tau)}{128 q^9}-\dfrac{63 \tau \cos (m q \tau)}{256 q^9}}{m^9} \\& +\dfrac{-\dfrac{13 \tau ^6 \sin (m q \tau )}{49152 q^{10}}-\dfrac{19 \tau ^4 \sin (m q \tau )}{768 q^{10}}-\dfrac{281 \tau ^2 \sin (m q \tau )}{1024 q^{10}}}{m^{10}} \\&
		+\dfrac{\dfrac{37 \tau ^7 \cos (m q \tau )}{1474560 q^{11}}+\dfrac{209 \tau ^5 \cos (m q \tau )}{49152 q^{11}}+\dfrac{263 \tau ^3 \cos (m q \tau )}{2048 q^{11}}+\dfrac{231 \tau  \cos (m q \tau )}{1024 q^{11}}}{m^{11}}
		+
		\mbox{ smaller terms}.
		\label{appneq24_4_full}
	\end{aligned}
\end{equation}
Now, the coefficient of the sine term in \eqref{appneq24_1_full} can be expanded for large $m$ as
\begin{equation}
	\dfrac{1}{m^2 q^2 \sqrt{1+m^2 q^2}}= \dfrac{1}{m^3 q^3}-\dfrac{1}{2 m^5 q^5}+\dfrac{3}{8 m^7 q^7}-\dfrac{5}{16 m^9 q^9}+\frac{35}{128 m^{11} q^{11}} +\mathcal{O}\left(m^{-12}\right).
	\label{appneq24_5_full}
\end{equation}
Using \eqref{appneq24_4_full} and \eqref{appneq24_5_full}, the expression in \eqref{appneq24_1_full} can be rewritten as
\begin{equation}
	\begin{aligned}
		\dfrac{1}{m^2 q^2 \sqrt{1+m^2 q^2}} &\sin \left(\dfrac{m q}{\sqrt{1+\dfrac{1}{m^2 q^2}}} \, \tau\right) = \dfrac{\sin (m q \tau)}{m^3 q^3} -\dfrac{\tau \cos (m q \tau)}{2 m^4 q^4}+\dfrac{-\dfrac{\tau^2 \sin (m q \tau)}{8 q^5}-\dfrac{\sin (m q \tau)}{2 q^5}}{m^5} \\  &+\dfrac{\dfrac{\tau^3 \cos (m q \tau)}{48 q^6}+\dfrac{5 \tau \cos (m q \tau)}{8 q^6}}{m^6}+\dfrac{\dfrac{\tau^4 \sin (m q \tau)}{384 q^7}+\dfrac{\tau^2 \sin (m q \tau)}{4 q^7}+\dfrac{3 \sin (m q \tau)}{8 q^7}}{m^7} \\ &+ \dfrac{-\dfrac{\tau^5 \cos (m q \tau)}{3840 q^8}-\dfrac{11 \tau^3 \cos (m q \tau)}{192 q^8}-\dfrac{11 \tau \cos (m q \tau)}{16 q^8}}{m^8} \\ &+\dfrac{-\dfrac{\tau^6 \sin (m q \tau)}{46080 q^9}-\dfrac{7 \tau^4 \sin (m q \tau)}{768 q^9}-\dfrac{47 \tau^2 \sin (m q \tau)}{128 q^9}-\dfrac{5 \sin (m q \tau)}{16 q^9}}{m^9} \\ & +\dfrac{\dfrac{\tau ^7 \cos (m q \tau )}{645120 q^{10}}+\dfrac{17 \tau ^5 \cos (m q \tau )}{15360 q^{10}}+\dfrac{27 \tau ^3 \cos (m q \tau )}{256 q^{10}}+\dfrac{93 \tau  \cos (m q \tau )}{128 q^{10}}}{m^{10}} \\ & +\dfrac{\dfrac{\tau ^6 \sin (m q \tau )}{9216 q^{11}}+\dfrac{31 \tau ^4 \sin (m q \tau )}{1536 q^{11}}+\dfrac{61 \tau ^2 \sin (m q \tau )}{128 q^{11}}+\dfrac{35 \sin (m q \tau )}{128 q^{11}}}{m^{11}} + \mbox{ smaller terms}.
	\end{aligned}
	\label{appneq24_6_full}
\end{equation}
Now, we put $m=2n-1$ back to Eq.\ \eqref{appneq24_6_full} and from Eqs.\ \eqref{appneq24_full} and \eqref{appneq24_1_full} we obtain
\begin{equation}
	\begin{aligned}
		S_2&=\sum_{n=N+1}^{\infty} \Bigg[ \dfrac{\sin ((2 n-1) q \tau )}{(2 n-1)^3 q^3}-\dfrac{\tau  \cos ((2 n-1) q \tau )}{2 (2 n-1)^4 q^4}+\dfrac{-\dfrac{\tau ^2 \sin ((2 n-1) q \tau )}{8 q^5}-\dfrac{\sin ((2 n-1) q \tau )}{2 q^5}}{(2 n-1)^5} \\ &+\dfrac{\dfrac{\tau ^3 \cos ((2 n-1) q \tau )}{48 q^6}+\dfrac{5 \tau  \cos ((2 n-1) q \tau )}{8 q^6}}{(2 n-1)^6}+\dfrac{\dfrac{\tau ^4 \sin ((2 n-1) q \tau )}{384 q^7}+\dfrac{\tau ^2 \sin ((2 n-1) q \tau )}{4 q^7}+\dfrac{3 \sin ((2 n-1) q \tau )}{8 q^7}}{(2 n-1)^7} \\ &+\dfrac{-\dfrac{\tau ^5 \cos ((2 n-1) q \tau )}{3840 q^8}-\dfrac{11 \tau ^3 \cos ((2 n-1) q \tau )}{192 q^8}-\dfrac{11 \tau  \cos ((2 n-1) q \tau )}{16 q^8}}{(2 n-1)^8} \\ &+\dfrac{-\dfrac{\tau ^6 \sin ((2 n-1) q \tau )}{46080 q^9}-\dfrac{7 \tau ^4 \sin ((2 n-1) q \tau )}{768 q^9}-\dfrac{47 \tau ^2 \sin ((2 n-1) q \tau )}{128 q^9}-\dfrac{5 \sin ((2 n-1) q \tau )}{16 q^9}}{(2 n-1)^9} \\& +\dfrac{\dfrac{\tau ^7 \cos ((2 n-1) q \tau )}{645120 q^{10}}+\dfrac{17 \tau ^5 \cos ((2 n-1) q \tau )}{15360 q^{10}}+\dfrac{27 \tau ^3 \cos ((2 n-1) q \tau )}{256 q^{10}}+\dfrac{93 \tau  \cos ((2 n-1) q \tau )}{128 q^{10}}}{(2 n-1)^{10}} \\ &+\dfrac{\dfrac{\tau ^6 \sin ((2 n-1) q \tau )}{9216 q^{11}}+\dfrac{31 \tau ^4 \sin ((2 n-1) q \tau )}{1536 q^{11}}+\dfrac{61 \tau ^2 \sin ((2 n-1) q \tau )}{128 q^{11}}+\dfrac{35 \sin ((2 n-1) q \tau )}{128 q^{11}}}{(2 n-1)^{11}} \\ &
		+ \mbox{smaller terms}\Bigg].
	\end{aligned}
	\label{appneq24_7_full}
\end{equation}
Evaluating the individual sums using Eq.\ \eqref{appneq12} and expanding for small $\tau$ gives
\begin{equation}
	\begin{aligned}
		S_2 &= \frac{\tau ^{3/2}}{4 q^2}+\left(-\frac{1}{8 q^2}-\frac{\pi }{8 q}\right) \tau ^2+\left(-\frac{1}{48 q^4}+\frac{1}{24 q^2}+\frac{1}{6}\right) \tau ^{5/2}+\left(\frac{1}{32 q^4}+\frac{1}{12}\right) \tau ^3 \\ &+\left(\frac{1}{320 q^6}-\frac{1}{48 q^4}-\frac{q^2}{90}+\frac{3}{40 q^2}\right) \tau ^{7/2}+\left(-\frac{1}{128 q^6}-\frac{q^2}{60}-\frac{1}{24 q^2}-\frac{5 \pi }{192 q}\right) \tau ^4 \\ &+\left(-\frac{1}{1792 q^8}+\frac{1}{128 q^6}+\frac{q^4}{1575}-\frac{q^2}{180}+\frac{5}{252 q^2}+\frac{1}{40}\right) \tau ^{9/2}+\left(\frac{1}{512 q^8}+\frac{q^4}{630}+\frac{1}{64 q^4}+\frac{1}{80}\right) \tau ^5 + \mathcal{O}\left(\tau^{11/2}\right)
	\end{aligned}
\end{equation}

\subsection{Expression for $S(\tau)$}
Finally, using Eq.\ \eqref{eq11}, all half-integer powers of $\tau$ cancel exactly, and we obtain
\begin{equation}
	\begin{aligned}
		S(\tau) &= \dfrac{2 \bar P}
		{\bar L} \Bigg[\dfrac{\pi  \tau  \tanh \left(\dfrac{\pi }{2 q}\right)}{4 q} -\dfrac{\pi  \tau ^2}{8 q} +  \left(\dfrac{\pi ^2 \text{sech}^2\left(\dfrac{\pi }{2 q}\right)}{96 q^2}+\dfrac{\pi  \tanh \left(\dfrac{\pi }{2 q}\right)}{12 q}-\dfrac{\pi  \sinh \left(\dfrac{\pi }{q}\right) \text{sech}^2\left(\dfrac{\pi }{2 q}\right)}{96 q}\right) \tau ^3 -\dfrac{5 \pi  \tau ^4}{192 q} \\ & + \left(-\dfrac{\pi ^3 \tanh \left(\dfrac{\pi }{2 q}\right) \text{sech}^2\left(\dfrac{\pi }{2 q}\right)}{7680 q^3}+\dfrac{13 \pi ^2 \text{sech}^2\left(\dfrac{\pi }{2 q}\right)}{7680 q^2}+\dfrac{17 \pi  \tanh \left(\dfrac{\pi }{2 q}\right)}{1280 q}-\dfrac{\pi  \sinh \left(\dfrac{\pi }{q}\right) \text{sech}^2\left(\dfrac{\pi }{2 q}\right)}{480 q}\right)\tau ^5 \\ &+ \mbox{smaller terms} \Bigg].
	\end{aligned}
	\label{appneq27_full}
\end{equation}
Substituting $q=\pi / \bar L$ in Eq.\ \eqref{appneq27_full} yields
\begin{equation}
	\begin{aligned}
		\bar u\left(\dfrac{\bar L}{2},\tau\right)=S(\tau)=& \bar P \Bigg[ \dfrac{\tau}{2}   \tanh \left(\frac{L}{2}\right) - \dfrac{\tau^2}{4} + \left\{\dfrac{1}{8} \tanh \left(\dfrac{L}{2}\right)+\dfrac{1}{48} L \text{sech}^2\left(\dfrac{L}{2}\right)\right\}\tau ^3 -\dfrac{5 \tau ^4}{96} \\ & +\left(-\dfrac{1}{480} L^2 \sinh ^4\left(\dfrac{L}{2}\right) \text{csch}^3(L)+\dfrac{7}{384} \tanh \left(\dfrac{L}{2}\right)+\dfrac{13 L \text{sech}^2\left(\dfrac{L}{2}\right)}{3840}\right)\tau ^5  + \cdots \Bigg].
	\end{aligned}
	\label{appneq28_full}
\end{equation}

\section{Comparison with Timoshenko beam model}
\label{timo}


\subsection{Timoshenko model: dimensional and nondimensional equations}
\label{timomodel}

In the Timoshenko model \cite{timoshenko1921}, the transverse displacement $u(x,t)$ and the rotation $\psi(x,t)$ of the cross-section are independent variables. The beam has length $L$, bending rigidity $EI$, mass per unit length $\rho A$, rotary inertia per unit length $\rho I$, and shear rigidity $\kappa_s G A$, where $G=E/[2(1+\nu)]$ is the shear modulus and $\kappa_s$ is the shear coefficient \cite{cowper1966shear}. A concentrated transverse impulse $P$ is applied at $t=0$ at $x=b$, and the beam starts from rest. The equations of motion are
\begin{equation}
	\begin{aligned}
		\rho A\,\dfrac{\partial^2 u}{\partial t^2}
		-\kappa_s G A\left(\dfrac{\partial^2 u}{\partial x^2}-\dfrac{\partial \psi}{\partial x}\right)
		&= P\,\delta(x-b)\,\delta(t),\\
		\rho I\,\dfrac{\partial^2 \psi}{\partial t^2}
		-EI\,\dfrac{\partial^2 \psi}{\partial x^2}
		-\kappa_s G A\left(\dfrac{\partial u}{\partial x}-\psi\right)
		&= 0 .
	\end{aligned}
	\label{eqtimodim}
\end{equation}
For a simply supported beam, the boundary conditions are zero displacement and zero bending moment at both ends,
\begin{equation}
	u=0, \qquad \dfrac{\partial \psi}{\partial x}=0 \qquad \text{at } x=0,\,L.
	\label{eqtimobc}
\end{equation}
We use the same scaling as for the Rayleigh beam, Eqs.\ \eqref{eqdim2} and \eqref{eq1nondscale}, namely
$\bar u = u\sqrt{A/I}$, $\bar x = x\sqrt{A/I}$, $\bar b = b\sqrt{A/I}$, $\bar L = L\sqrt{A/I}$, $\bar P = P/\sqrt{EA\rho I}$, $\tau = t\sqrt{EA/(\rho I)}$ and $\bar v=\partial \bar u/\partial \tau= v \sqrt{\rho / E}$. The rotation $\psi$ is already dimensionless and is not scaled. Equations \eqref{eqtimodim} become
\begin{equation}
	\begin{aligned}
		\dfrac{\partial^2 \bar u}{\partial \tau^2}
		&= g\left(\dfrac{\partial^2 \bar u}{\partial \bar x^2}-\dfrac{\partial \psi}{\partial \bar x}\right)
		+\bar P\,\delta(\bar x-\bar b)\,\delta(\tau),\\
		\dfrac{\partial^2 \psi}{\partial \tau^2}
		&= \dfrac{\partial^2 \psi}{\partial \bar x^2}
		+g\left(\dfrac{\partial \bar u}{\partial \bar x}-\psi\right),
	\end{aligned}
	\label{eqtimo}
\end{equation}
with the single dimensionless parameter
\begin{equation}
	g=\dfrac{\kappa_s G}{E}=\dfrac{\kappa_s}{2(1+\nu)}.
	\label{eqg}
\end{equation}
The parameter $g$ is the square of the ratio of the shear-wave speed $\sqrt{\kappa_s G/\rho}$ to the scaling speed $\sqrt{E/\rho}$. It is $1/\gamma^2$ in the notation of Han {\it et al.}\ \cite{han1999dynamics}. For a rectangular section, $\nu=0.3$ and $\kappa_s=0.85$ \cite{cowper1966shear} give $g=0.327$. For physical values $\kappa_s\le 1$ and $\nu\ge 0$, so $0<g\le 1/2$, which is assumed throughout. The boundary conditions \eqref{eqtimobc} become
\begin{equation}
	\bar u=0,\qquad \dfrac{\partial \psi}{\partial \bar x}=0 \qquad \text{at } \bar x=0,\,\bar L.
	\label{eqtimobcnd}
\end{equation}
The Rayleigh beam is the $g\to\infty$ limit of Eq.\ \eqref{eqtimo}. To see this, differentiate the second of Eqs.\ \eqref{eqtimo} with respect to $\bar x$ and use the first,
\begin{equation}
	\dfrac{\partial^3 \psi}{\partial \bar x\,\partial\tau^2}-\dfrac{\partial^3 \psi}{\partial \bar x^3}
	=\dfrac{\partial^2 \bar u}{\partial \tau^2}-\bar P\delta(\bar x-\bar b)\delta(\tau).
\end{equation}
As $g\to\infty$ the shear strain $\partial\bar u/\partial \bar x-\psi$ vanishes while the shear force $g(\partial\bar u/\partial \bar x-\psi)$ stays finite, so $\psi\to\partial \bar u/\partial \bar x$, and the equation above reduces to Eq.\ \eqref{eq1nond}, thus recovering the Rayleigh beam model.

Substituting $\bar u=\bar u_0 \exp[i(\kappa\bar x-\omega\tau)]$ and $\psi= \psi_0 \exp[i(\kappa\bar x-\omega\tau)]$ into the homogeneous form of Eq.\ \eqref{eqtimo} gives the dispersion relation
\begin{equation}
	\omega^4-\omega^2\left[(1+g)\kappa^2+g\right]+g\kappa^4=0 ,
	\label{eqdisptimo}
\end{equation}
which is the characteristic equation of Han {\it et al.}\ \cite{han1999dynamics} in the present scaling. The two branches are
\begin{equation}
	\omega_\pm^2(\kappa)=\tfrac12\left[S(\kappa)\pm\Delta(\kappa)\right],\quad
	S=(1+g)\kappa^2+g,\quad
	\Delta=\sqrt{S^2-4g\kappa^4}=\sqrt{(1-g)^2\kappa^4+2g(1+g)\kappa^2+g^2}.
	\label{eqbranches}
\end{equation}
Since $\Delta>0$ for all $\kappa$, the branches do not cross. The upper branch has the cutoff $\omega_+(0)=\sqrt g$, which is the critical frequency $\sqrt{\kappa_sGA/\rho I}$ of Han {\it et al.}\ \cite{han1999dynamics}. At large $\kappa$, $\omega_-\to\sqrt g\,\kappa$ (shear waves) and $\omega_+\to\kappa$. The upper branch is the `second spectrum' of the Timoshenko beam, whose physical interpretation has been debated \cite{traillnash1953effects,stephen2006second,levinson1982}.


\subsection{Finite simply supported Timoshenko beam}
\label{timofinite}

Modal expansion for the forced Timoshenko beam follows Dolph \cite{dolph1954}, Herrmann \cite{herrmann1955} and Han {\it et al.}\ \cite[Sec.~5]{han1999dynamics}. For the simply supported beam, the conditions \eqref{eqtimobcnd} are satisfied by
\begin{equation}
	\bar u(\bar x,\tau)=\sum_{k=1}^{\infty}a_k(\tau)\sin(p_k\bar x),\qquad
	\psi(\bar x,\tau)=\sum_{k=1}^{\infty}c_k(\tau)\cos(p_k\bar x),\qquad p_k=\dfrac{k\pi}{\bar L},
	\label{eqtimomod1}
\end{equation}
with $a_k(0)=c_k(0)=0$ and $a_k'(0^-)=c_k'(0^-)=0$. Substituting the modal expansions of Eq.\ \eqref{eqtimomod1} into Eq.\ \eqref{eqtimo}, then using the orthogonality of the sine and cosine functions, gives a coupled pair of ordinary differential equations for each mode. In matrix form, with $\mathbf q_n=(a_n,c_n)^T$ and primes denoting derivatives with respect to $\tau$,
\begin{equation}
	\mathbf q_n''+\mathbf K_n\mathbf q_n=\dfrac{2\bar P\sin(p_n\bar b)}{\bar L}\begin{pmatrix}1\\0\end{pmatrix}\delta(\tau),
	\qquad
	\mathbf K_n=\begin{pmatrix} g p_n^2 & -g p_n\\ -g p_n & p_n^2+g\end{pmatrix}.
	\label{eqtimomod2}
\end{equation}
The eigenvalues of $\mathbf K_n$ are $\omega_{n\pm}^2=\omega_\pm^2(p_n)$, as given by Eq.\ \eqref{eqbranches}, with $\omega_{n+}^2-\omega_{n-}^2=\Delta_n$. The corresponding modal weights in the displacement response are
\begin{equation}
	w_{n\pm}=\dfrac{g p_n^2-\omega_{n\mp}^2}{\omega_{n\pm}^2-\omega_{n\mp}^2}
	=\dfrac12\mp\dfrac{(1-g)p_n^2+g}{2\Delta_n},
	\qquad w_{n+}+w_{n-}=1.
	\label{eqweights}
\end{equation}
Projecting Eq.\ \eqref{eqtimomod2} onto the two eigenmodes and applying the impulse gives
\begin{equation}
	a_n(\tau)=\dfrac{2\bar P\sin(p_n\bar b)}{\bar L}\sum_{\pm}w_{n\pm}\dfrac{\sin(\omega_{n\pm}\tau)}{\omega_{n\pm}},\qquad
	c_n(\tau)=\dfrac{2\bar P\sin(p_n\bar b)}{\bar L}\,\dfrac{g p_n}{\Delta_n}\left[\dfrac{\sin(\omega_{n-}\tau)}{\omega_{n-}}-\dfrac{\sin(\omega_{n+}\tau)}{\omega_{n+}}\right].
	\label{eqtimomod4}
\end{equation}
Substituting Eq.\ \eqref{eqtimomod4} into the modal expansions in Eq.\ \eqref{eqtimomod1} gives
\begin{align}
	&\bar u(\bar x,\tau)=\dfrac{2\bar P}{\bar L}\sum_{n=1}^{\infty}\sin(p_n\bar b)\sin(p_n\bar x)\sum_{\pm}w_{n\pm}\dfrac{\sin(\omega_{n\pm}\tau)}{\omega_{n\pm}},\label{eqtimomod7} \\
	&\psi(\bar x,\tau)=\dfrac{2\bar P}{\bar L}\sum_{n=1}^{\infty} \dfrac{\sin(p_n\bar b)\, g p_n}{\Delta_n}\left[\dfrac{\sin(\omega_{n-}\tau)}{\omega_{n-}}-\dfrac{\sin(\omega_{n+}\tau)}{\omega_{n+}}\right]\cos(p_n\bar x).
	\label{eqtimomod7cn}
\end{align}
Equation \eqref{eqtimomod7} is the exact impulse response of the finite simply supported Timoshenko beam for an impulse at $\bar x=\bar b$, observed at any spatial location $0\leq \bar x \leq \bar L$.

We now consider the response at the point of application, $\bar x=\bar b=\bar L/2$. Only odd modes survive. Writing $q=\pi/\bar L$, so that $p_n=(2n-1)q$, Eq.\ \eqref{eqtimomod7} gives
\begin{equation}
	\bar u\!\left(\dfrac{\bar L}{2},\tau\right)=\dfrac{2\bar P}{\bar L}\sum_{n=1}^{\infty}\sum_{\pm}w_{n\pm}\,\dfrac{\sin(\omega_{n\pm}\tau)}{\omega_{n\pm}},
	\label{eqtimomid1}
\end{equation}
where
\begin{equation}
	\omega_{n\pm}^2=\tfrac12\Big[(1+g)p_n^2+g\pm\Delta_n\Big],\quad
	\Delta_n=\sqrt{(1-g)^2p_n^4+2g(1+g)p_n^2+g^2},\quad
	w_{n\pm}=\dfrac12\mp\dfrac{(1-g)p_n^2+g}{2\Delta_n}.
	\label{eqtimomid2}
\end{equation}
Differentiating with respect to $\tau$,
\begin{equation}
	\bar v\!\left(\dfrac{\bar L}{2},\tau\right)=\dfrac{2\bar P}{\bar L}\sum_{n=1}^{\infty}\sum_{\pm}w_{n\pm}\cos(\omega_{n\pm}\tau).
	\label{eqtimomid3}
\end{equation}
Equations \eqref{eqtimomid1}-\eqref{eqtimomid3} are the displacement and velocity responses, respectively, of finite Timoshenko beam, and are the counterparts of Eqs.\ \eqref{eqmod8} and \eqref{eqvel}.


\subsection{Infinite Timoshenko beam}
\label{timoinf}

The infinite beam response is obtained directly from Eq.\ \eqref{eqtimo}, in the same way as for the Rayleigh beam. With the impulse applied at the origin, $\bar b=0$, let
\begin{equation}
	\hat u(\kappa,\tau)=\int_{-\infty}^{\infty}\bar u(\bar x,\tau)\,e^{-i\kappa\bar x}\,{\rm d}\bar x,\qquad
	\hat\psi(\kappa,\tau)=\int_{-\infty}^{\infty}\psi(\bar x,\tau)\,e^{-i\kappa\bar x}\,{\rm d}\bar x .
\end{equation}
The displacement $\bar u$ is even in $\bar x$ and the rotation $\psi$ is odd, so $\hat\psi$ is purely imaginary. We therefore write $\hat\psi=i\hat\chi$ with $\hat\chi$ real. Transforming Eq.\ \eqref{eqtimo} gives
\begin{equation}
	\begin{aligned}
		\dfrac{\partial^2\hat u}{\partial\tau^2}+g\kappa^2\hat u-g\kappa\hat\chi&=\bar P\delta(\tau),\\
		\dfrac{\partial^2\hat \chi}{\partial\tau^2}+(\kappa^2+g)\hat\chi-g\kappa\hat u&=0 .
	\end{aligned}
	\label{Teqfs1}
\end{equation}
This is the system \eqref{eqtimomod2} with $p_n$ replaced by the continuous wavenumber $\kappa$. Eliminating $\hat\chi$ gives a single fourth-order equation in $\tau$,
\begin{equation}
	\dfrac{\partial^4\hat u}{\partial\tau^4}+\left[(1+g)\kappa^2+g\right]\dfrac{\partial^2\hat u}{\partial\tau^2}+g\kappa^4\hat u
	=\bar P\left[\delta''(\tau)+(\kappa^2+g)\delta(\tau)\right],
	\label{Teqfs2}
\end{equation}
Its characteristic polynomial is the dispersion relation \eqref{eqdisptimo}. The system \eqref{Teqfs1} is solved with a Laplace transform in $\tau$ (Laplace variable $s$, zero initial condition). Writing $\tilde u$ and $\tilde\chi$ for the transforms,
\begin{equation}
	(s^2+g\kappa^2)\,\tilde u-g\kappa\tilde\chi=\bar P,\qquad (s^2+\kappa^2+g)\tilde\chi=g\kappa\tilde u,
\end{equation}
so that
\begin{equation}
	\tilde u(\kappa,s)=\bar P\,\dfrac{s^2+\kappa^2+g}{s^4+\left[(1+g)\kappa^2+g\right]s^2+g\kappa^4}
	=\bar P\left[\dfrac{w_+}{s^2+\omega_+^2}+\dfrac{w_-}{s^2+\omega_-^2}\right],
	\label{Teqfs3}
\end{equation}
where $\omega_\pm(\kappa)$ are given by Eq.\ \eqref{eqbranches} and $w_\pm(\kappa)$ by Eq.\ \eqref{eqweights} with $p_n\to\kappa$. Inverting, and treating $\hat\chi$ in the same way,
\begin{equation}
	\hat u(\kappa,\tau)=\bar P\sum_{\pm}w_\pm\dfrac{\sin(\omega_\pm\tau)}{\omega_\pm},\qquad
	\hat \chi(\kappa,\tau)=\bar P\dfrac{g\kappa}{\Delta}\left[\dfrac{\sin(\omega_-\tau)}{\omega_-}-\dfrac{\sin(\omega_+\tau)}{\omega_+}\right].
	\label{Teqfs4}
\end{equation}
Since $\hat u$ is even in $\kappa$, the inverse Fourier transform is
\begin{equation}
	\bar u_\infty(\bar x,\tau)=\dfrac1\pi\int_0^\infty\hat u(\kappa,\tau)\cos(\kappa\bar x)\,{\rm d}\kappa
	=\dfrac{\bar P}{\pi}\int_0^\infty\sum_{\pm}w_\pm(\kappa)\dfrac{\sin(\omega_\pm(\kappa)\tau)}{\omega_\pm(\kappa)}\cos(\kappa\bar x)\,{\rm d}\kappa .
	\label{Teqinfx}
\end{equation}
Equation~\eqref{Teqinfx} is the impulse response of infinite Timoshenko beam at any spatial location $\bar x$ away from the struck point, expressed in continuum spectral integral form. The two branches correspond to the lower and upper spectra of the Timoshenko beam. Differentiating with respect to $\tau$ gives the velocity response,
\begin{equation}
	\bar v_\infty(\bar x,\tau)=\dfrac{\partial\bar u_\infty}{\partial\tau}
	=\dfrac{\bar P}{\pi}\int_0^\infty\sum_{\pm}w_\pm(\kappa)\cos(\omega_\pm(\kappa)\tau)\cos(\kappa\bar x)\,{\rm d}\kappa .
	\label{Teqinfv}
\end{equation}
Now we consider the struck-point response by putting $\bar x=0$, yielding
\begin{equation}
	\bar u_\infty(0,\tau)=\dfrac{\bar P}{\pi}\int_0^\infty\sum_{\pm}w_\pm\dfrac{\sin(\omega_\pm\tau)}{\omega_\pm}\,{\rm d}\kappa,\qquad
	\bar v_\infty(0,\tau)=\dfrac{\bar P}{\pi}\int_0^\infty\sum_{\pm}w_\pm\cos(\omega_\pm\tau)\,{\rm d}\kappa .
	\label{Teqsecn10}
\end{equation}
Unlike the Rayleigh beam, the Timoshenko point response has an initial displacement jump $\bar P/(2\sqrt g)$ and a corresponding singular velocity contribution $\bar P\delta(\tau)/(2\sqrt g)$, both of which vanish as $g\to\infty$, recovering the Rayleigh beam behaviour.

Similar to the methodology adapted for the Rayleigh beam, a closed form of Eq.\ \eqref{Teqsecn10} is obtained in the Laplace domain. Taking the Laplace transform of Eq.\ \eqref{Teqsecn10}$_1$, interchanging the order of integration and using Eq.\ \eqref{Teqfs3},
\begin{equation}
	\tilde u_\infty(0,s)=\dfrac{\bar P}{\pi}\int_0^\infty\dfrac{s^2+\kappa^2+g}{g\kappa^4+(1+g)s^2\kappa^2+s^2(s^2+g)}\,{\rm d}\kappa .
	\label{Teqlap1}
\end{equation}
For ${\rm Re}\,s>0$, the denominator factors as $g(\kappa^2+a^2)(\kappa^2+b^2)$ with ${\rm Re}\,a,{\rm Re}\,b>0$ and
\begin{equation}
	a^2+b^2=\dfrac{(1+g)s^2}{g},\qquad a^2b^2=\dfrac{s^2(s^2+g)}{g},\qquad\text{so}\quad ab=\dfrac{s\sqrt{s^2+g}}{\sqrt g}.
\end{equation}
The standard integral $\displaystyle \int_0^\infty(\kappa^2+c)\,{\rm d}\kappa/[(\kappa^2+a^2)(\kappa^2+b^2)]=\pi(c+ab)/[2ab(a+b)]$ with $c=s^2+g$ gives
\begin{equation}
	\tilde u_\infty(0,s)=\dfrac{\bar P}{2g}\,\dfrac{s^2+g+ab}{ab\,(a+b)}.
\end{equation}
Now $g(a+b)^2=(1+g)s^2+2\sqrt g\,s\sqrt{s^2+g}=Z^2-g^2$, where
\begin{equation}
	Z(s)=s+\sqrt{g(s^2+g)} .
\end{equation}
Simplifying,
\begin{equation}
	\tilde u_\infty(0,s)=\dfrac{\bar P}{2\sqrt g}\,\dfrac{Z}{s\sqrt{Z^2-g^2}}
	=\dfrac{\bar P}{2\sqrt g}\,\dfrac{s+\sqrt{g(s^2+g)}}{s^{3/2}\sqrt{(1+g)s+2\sqrt{g(s^2+g)}}},
	\label{Teqlap2}
\end{equation}
and the transform of the velocity is $s\tilde u_\infty(0,s)$. The inverse Laplace transform of Eq.\ \eqref{Teqlap2} is not elementary and is evaluated numerically. In the limit $g\to\infty$, $Z= g+s+s^2/2$ and $Z^2-g^2=gs(s+2)$, giving
\begin{equation}
	\tilde u_\infty(0,s)\to\dfrac{\bar P}{2s\sqrt{s(s+2)}},
\end{equation}
which recovers the Rayleigh result in Eq.\ \eqref{eqsecn14}.

\subsection{Short- and long-time asymptotics of infinite Timoshenko beam}
\label{timoasymp}

Following similar method of Section~\ref{contspecasym}, now derive the short- and long-time asymptotics of the impulse response of infinite Timoshenko beam. The symbolic computations are done in Mathematica.

Large-$s$ expansion of Eq.\ \eqref{Teqlap2} yields
\begin{equation}
	\begin{aligned}
		\tilde u_\infty(0,s)_{\rm asymp}&=\frac {P} {2\sqrt {g} s} + \frac {g^{3/2} P} {4\left (\sqrt {g} + 
			1 \right)^2 s^3} - \frac {\left (\sqrt {g} + 
			4 \right) g^3 P} {16\left (\sqrt {g} + 
			1 \right)^4 s^5} + \frac {g^4\left (g^{3/2} + 6 g + 
			12\sqrt {g} + 
			2 \right) P} {32\left (\sqrt {g} + 
			1 \right)^6 s^7}  \\ & -\frac {g^5\left (5 g^{5/2} + 128 g^{3/2} + 
			40 g^2 + 184 g + 64\sqrt {g} + 
			8 \right) P} {256\left (\sqrt {g} + 1 \right)^8 s^9} + \ldots .
	\end{aligned}
	\label{timasympshorts}
\end{equation}
Applying inverse Laplace transform term by term on Eq.\ \eqref{timasympshorts} yields
\begin{equation}
	\begin{aligned}
		\bar u_\infty(0,\tau)_{\rm asymp}=\frac {P} {2\sqrt {g}}\, & \Biggg( 1+ \frac {g^2\tau^2} {4\left (\sqrt {g} + 1 \right)^2} - \frac {\left (4 g^{7/2} + g^4 \right)\tau^4} {192\left (\sqrt {g} + 
			1 \right)^4}+\frac {\left (g^{3/2} + 6 g + 12\sqrt{g} + 2 \right) g^{9/2}\tau^6} {11520\left (\sqrt {g} + 1 \right)^6} \\ & - \frac {\left (128 g^{3/2} + 5 g^{5/2} + 
			40 g^2 + 184 g + 64\sqrt {g} + 8 \right) g^{11/2}\tau^8} {5160960\left (\sqrt {g} + 1 \right)^8} + \mathcal{O}(\tau^{10})\Biggg)
	\end{aligned}
	\label{timasympshorttau}
\end{equation}
It is interesting to note, that this trick of using Laplace transform to find the short-time response works only for finite $g$ held fixed as we consider small $\tau$. If we hold this series fixed and make $g\rightarrow \infty$, this approximation breaks down. A more nuanced analysis is not attempted as the Timoshenko beam is not our primary subject of study in this paper.

Similarly, small-$s$ expansion of Eq.\ \eqref{Teqlap2} yields
\begin{equation}
	\begin{aligned}
		\tilde u_\infty(0,s)_{\rm asymp, \infty}&=\dfrac{P}{2\sqrt{2}\,s^{3/2}}
		-\dfrac{(-3+g)P}{8\sqrt{2}\,g\sqrt{s}}
		+\dfrac{1}{2\sqrt{g}}
		\left(
		\dfrac{1}{2\sqrt{2}\sqrt{g}}
		-\dfrac{1+g}{4\sqrt{2}\,g^{3/2}}
		+\dfrac{3-2g+3g^2}{32\sqrt{2}\,g^{3/2}}
		\right)P\sqrt{s} \\ &
		+\dfrac{1}{2\sqrt{g}}
		\left(
		-\dfrac{1+g}{8\sqrt{2}\,g^{3/2}}
		+\dfrac{3-2g+3g^2}{32\sqrt{2}\,g^{5/2}}
		-\dfrac{(1+g)(5-14g+5g^2)}{128\sqrt{2}\,g^{5/2}}
		\right)Ps^{3/2} \\ &
		+\dfrac{1}{2\sqrt{g}}
		\Biggg(
		-\dfrac{1}{8\sqrt{2}\,g^{3/2}}
		+\dfrac{3-2g+3g^2}{64\sqrt{2}\,g^{5/2}}
		-\dfrac{(1+g)(5-14g+5g^2)}{128\sqrt{2}\,g^{7/2}} \\ &
		+\dfrac{5(7-20g+10g^2-20g^3+7g^4)}{2048\sqrt{2}\,g^{7/2}}
		\Biggg)Ps^{5/2} + \ldots.
	\end{aligned}
	\label{timasymplongs}
\end{equation}
Applying inverse Laplace transform term by term on Eq.\ \eqref{timasymplongs} yields
\begin{equation}
	\begin{aligned}
		\bar u_\infty(0,\tau)_{\rm asymp,\infty}=\dfrac {P\sqrt {\tau}} {\sqrt {2\pi}}\,  &\Biggg(1+ \dfrac {\dfrac {3} {8 g} - \dfrac {1} {8}} {\tau} + \dfrac {\dfrac {5} {128 g^2} - \dfrac {3} {64 g} - \dfrac {3} {128}} {\tau^2} + \dfrac {\dfrac {21} {1024 g^3} - \dfrac {45} {1024 g^2} +
			\dfrac {15} {1024 g} - \dfrac {15} {1024}} {\tau^3} \\& + \dfrac {\dfrac {675} {32768 g^4} - \dfrac {525} {8192 g^3} \
			+ \dfrac {945} {16384 g^2} + \dfrac {315} {8192 g} - \dfrac {525} {32768}}
		{\tau^4} +\mathcal{O}(\tau^{-5}) \Biggg).
	\end{aligned}
	\label{timasymplongtau}
\end{equation}
It is interesting to note from these two asymptotics is that the short-time behaviour is different than EB and Rayleigh beam, but in the long-time, EB, Rayleigh and Timoshenko model all show exactly same $\sqrt{\tau}$ behaviour: see Eqs.\ \eqref{eqlongcont} and \eqref{timasymplongtau}.

\subsection{Numerical results}
\label{timocomp}

The preceding sections in the main text establish an analytical expression for the struck-point displacement and velocity of an infinite Rayleigh beam and identify when the response of a finite beam begins to depart from it. This section examines whether the same approximation is useful when shear deformation is included. We compare the finite and infinite responses predicted by the Rayleigh and Timoshenko models, using the Timoshenko formulation worked out for our boundary conditions in Appendix~\ref{timo}. The comparison first examines the onset of finite-boundary effects in each model and then evaluates the differences between the infinite Rayleigh response and the finite Rayleigh, finite Timoshenko, and infinite Timoshenko responses up to the Rayleigh onset time.

The Timoshenko model includes shear deformation and rotary inertia, whereas the Rayleigh model includes rotary inertia but neglects shear deformation. Their different dispersion relations produce different transient responses. The comparison is therefore not intended to establish that one model is generally more accurate than the other. It tests whether the infinite Rayleigh solution provides a useful approximation to the finite beam response over a specified time interval.

However, one feature of the Timoshenko model deserves comment. Its impulse response predicts an instantaneously nonzero finite displacement at the struck point. Presumably this is because, within the Timoshenko framework, a very narrow section of the beam can undergo a large shear. In this sense, a Timoshenko beam can be cut by a localized impulse, whereas a Rayleigh beam cannot. Which behaviour is physically more appropriate may be debated. Of course, real impact loads build up over a finite time, are bounded, and act over a nonzero region, so in such physical problems the instantaneously nonzero displacement in the point impulse response may not produce net unphysical effects. In any case, although we use the Timoshenko model as the reference when assessing the infinite Rayleigh approximation, we do have some reservations about the instantaneously nonzero response of the Timoshenko beam.

\begin{figure}[htbp]
	\centering
	\includegraphics[width=0.95\textwidth]{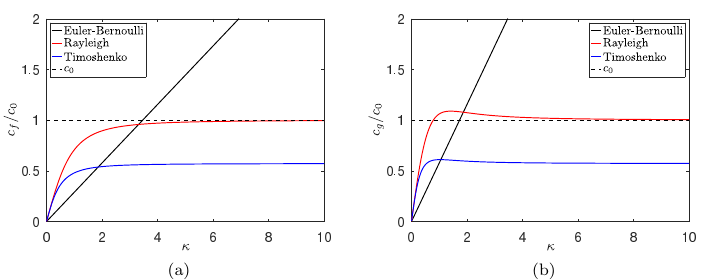}
	\caption{Phase velocity $c_f$ (left) and group velocity $c_g$ (right) of flexural waves in Euler-Bernoulli (black), Rayleigh (red) and Timoshenko (blue, $\nu=0.3$, shear coefficient $0.85$) beams; normalized by $c_0=\sqrt{E/\rho}$ (see Eq.~\eqref{eq1nondscale}).}
	\label{fig:groupvel3}
\end{figure}

Figure~\ref{fig:groupvel3} compares the phase velocity $c_f$ and group velocity $c_g$ of the three models, normalized by $c_0=\sqrt{E/\rho}$ (recall Eq.~\eqref{eq1nondscale}). Only the lower, flexural branch of the Timoshenko model is shown, as it is the branch relevant to the transverse flexural response considered here. The phase and group velocities of EB beam shown here grow without bound, and are clearly unphysical. For the parameters used, the phase and group velocities of the Timoshenko flexural branch both tend to $\sqrt{g}\approx0.57$ at high wavenumbers, the former from below and the latter from above. The group velocity reaches a maximum of about $0.61$ at $\kappa_*\approx1.03$. In contrast, for the Rayleigh beam, the phase and group velocities tend to unity at high wavenumbers, and its group velocity reaches a maximum of about $1.089$ at $\kappa_*=\sqrt{2}$. Shear deformation therefore lowers both the high-wavenumber limit and the peak group velocity, and shifts the peak to a smaller wavenumber. Bounded phase and group velocities do not, however, imply a finite signal (as is well known; see \cite{brillouin1960wave}).

The wavelength corresponding to $\kappa_*=\sqrt{2}$ is $2\pi/(\sqrt{2}\sqrt{A/I})\approx4.4\sqrt{I/A}$. For a square section of side $h$, this is about $1.3h$. At such short wavelengths, one-dimensional beam theories are physically inapplicable \cite{zak2012assessment}, although mathematical solutions to the governing PDEs remain of academic interest. The results near the group velocity maximum should therefore be interpreted within the limitations of the Rayleigh beam model, when physical behaviour is of interest.

\begin{figure}[htbp]
	\centering
	\includegraphics[width=0.95\textwidth]{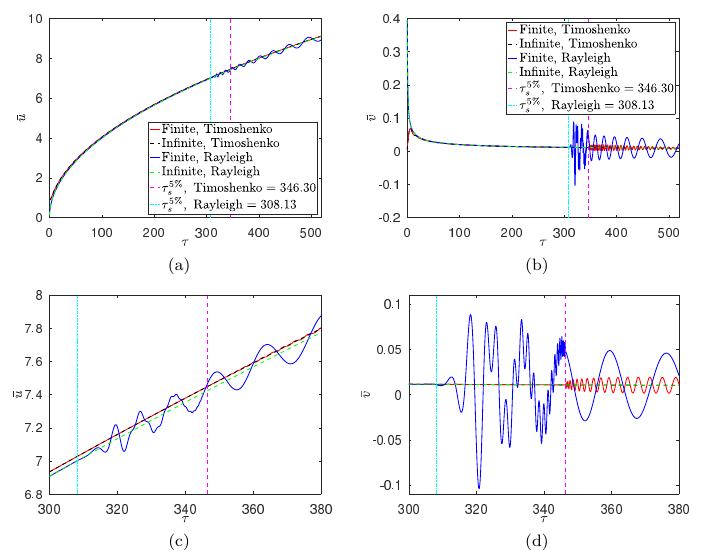}
	\caption{Struck-point displacement and velocity responses for finite and infinite Rayleigh and Timoshenko beams with $\bar L=346.4$. Panels (a) and (b) show displacement and velocity, respectively, and panels (c) and (d) show enlarged views. The markers indicate the 5\% onset times for the finite Rayleigh and Timoshenko beams.}
	\label{figstruck3}
\end{figure}

Figure~\ref{figstruck3} compares the four responses for $\bar L=346.4$ (which corresponds to a 1 m long steel beam, with a 1 cm $\times$ 1 cm cross section, as used in the example in Section~\ref{validation}). The finite and infinite solutions of each model agree initially and subsequently separate as the effects of the finite supports become significant. The Rayleigh and Timoshenko responses also differ from each other in ways that include a displacement jump and a singular velocity contribution at the point of impact, as discussed above.

For this beam length, the 5\% onset (as defined in Section~\ref{validation}) occurs at $\tau_s^{5\%}=308.13$ for the Rayleigh model and at $\tau_s^{5\%}=346.30$ for the Timoshenko model. This is because the local maximum in group velocity is relatively stronger for the Rayleigh beam, while it is smaller for the Timoshenko beam (recall Fig.~\ref{fig:groupvel}(b)).


\section{Derivation of finite-length correction}
\label{finlender}

Consider the modal solution of Eq.\ \eqref{eqmod7} for a Rayleigh beam of length $\bar L$, impulse applied at $\bar x=\bar b$, and response observed at $\bar x$, with $\bar p_n=n\pi/\bar L= \kappa $. Using the product-to-sum identity for $\sin(\bar p_n\bar b)\sin(\bar p_n\bar x)$, the response can be written as
\begin{equation}
	\bar u_F(\bar x,\tau)=\dfrac{2\bar P}{\bar L}\sum_{n=1}^{\infty}f(nq,\tau), \qquad q=\dfrac{\pi}{\bar L},
	\label{eqimg1}
\end{equation}
where
\begin{equation}
	f( \kappa ,\tau)=\dfrac{\left[\cos\left( \kappa (\bar x-\bar b)\right)-\cos\left( \kappa (\bar x+\bar b)\right)\right]\sin\left(\dfrac{ \kappa ^2}{\sqrt{1+ \kappa ^2}}\tau\right)}{2 \kappa ^2\sqrt{1+ \kappa ^2}}.
	\label{eqimg2}
\end{equation}
Note that in this section we are denoting the finite beam impulse response as $\bar u_F$ and the infinite beam impulse response as $\bar u_\infty$. The function $f( \kappa ,\tau)$ is even in $ \kappa $, and a short-$ \kappa $ expansion shows $f( \kappa ,\tau)=\bar x\bar b\,\tau\, \kappa ^2+O( \kappa ^4)$, so that $f(0,\tau)=0$; consequently the sum over positive integers can be extended to the full integer lattice,
\begin{equation}
	\bar u_F(\bar x,\tau)=\dfrac{2\bar P}{\bar L}\sum_{n=1}^{\infty}f(nq,\tau)=\dfrac{\bar P}{\bar L}\sum_{n=-\infty}^{\infty}f(nq,\tau).
	\label{eqimg3}
\end{equation}

We now evaluate the lattice sum $\sum_n f(nq,\tau)$ using the Poisson summation formula, which relates a sum of a function over an equally spaced lattice to a sum of its Fourier transform over the reciprocal lattice. For a sufficiently well-behaved function $g( \kappa )$ and lattice spacing $q$,
\begin{equation}
	\sum_{n=-\infty}^{\infty}g(nq)=\dfrac{1}{q}\sum_{j=-\infty}^{\infty}\hat g\left(\dfrac{2\pi j}{q}\right), \qquad
	\hat g(\omega)=\int_{-\infty}^{\infty}g( \kappa )\,e^{-i\omega  \kappa }\,{\rm d} \kappa .
	\label{eqimgpoisson}
\end{equation}
Taking $g( \kappa )=f( \kappa ,\tau)$ and using $q=\pi/\bar L$, the reciprocal-lattice spacing becomes
\begin{equation}
	\dfrac{2\pi j}{q}=2j\bar L.
	\label{eqimgrecip}
\end{equation}
Thus Eq.\ \eqref{eqimgpoisson} gives
\begin{equation}
	\sum_{n=-\infty}^{\infty}f(nq,\tau)=\dfrac{1}{q}\sum_{j=-\infty}^{\infty}\hat f(2j\bar L,\tau), \qquad
	\hat f(\omega,\tau)=\int_{-\infty}^{\infty}f( \kappa ,\tau)\,e^{-i\omega  \kappa }\,{\rm d} \kappa .
	\label{eqimgfourier}
\end{equation}
Since $f( \kappa ,\tau)$ is even and real, its Fourier transform reduces to a cosine integral,
\begin{equation}
	\hat f(\omega,\tau)=2\int_0^\infty f( \kappa ,\tau)\cos(\omega  \kappa )\,{\rm d} \kappa ,
	\label{eqimgcosine}
\end{equation}
and $\hat f(\omega,\tau)$ is itself even in $\omega$. Separating the $j=0$ term from the $j$, $-j$ pairs in Eq.\ \eqref{eqimgfourier}, and using Eq.\ \eqref{eqimgcosine} together with $1/q=\bar L/\pi$, gives
\begin{equation}
	\sum_{n=-\infty}^{\infty}f(nq,\tau)
	=\dfrac{2\bar L}{\pi}\int_0^\infty f( \kappa ,\tau)\,{\rm d} \kappa 
	+\dfrac{4\bar L}{\pi}\sum_{j=1}^{\infty}\int_0^\infty f( \kappa ,\tau)\cos(2j\bar L \kappa )\,{\rm d} \kappa .
	\label{eqimg4}
\end{equation}
Substituting Eq.\ \eqref{eqimg4} into Eq.\ \eqref{eqimg3} then yields
\begin{equation}
	\bar u_F(\bar x,\tau)=\dfrac{2\bar P}{\pi}\int_0^\infty f( \kappa ,\tau)\,{\rm d} \kappa 
	+\dfrac{4\bar P}{\pi}\sum_{j=1}^{\infty}\int_0^\infty f( \kappa ,\tau)\cos(2j\bar L \kappa )\,{\rm d} \kappa .
	\label{eqimg4b}
\end{equation}

We now interpret each term of Eq.\ \eqref{eqimg4b} using the definition of $f( \kappa ,\tau)$ in Eq.\ \eqref{eqimg2} and the infinite beam response of Eq.\ \eqref{eqinfx}. The first term of Eq.\ \eqref{eqimg4b} is immediate, since it involves $f( \kappa ,\tau)$ directly,
\begin{equation}
	\dfrac{2\bar P}{\pi}\int_0^\infty f( \kappa ,\tau)\,{\rm d} \kappa 
	=\dfrac{\bar P}{\pi}\int_0^\infty
	\dfrac{\left[\cos\left( \kappa (\bar x-\bar b)\right)-\cos\left( \kappa (\bar x+\bar b)\right)\right]\sin\left(\dfrac{ \kappa ^2}{\sqrt{1+ \kappa ^2}}\tau\right)}{ \kappa ^2\sqrt{1+ \kappa ^2}}\,{\rm d} \kappa 
	=\bar u_\infty(\bar x-\bar b,\tau)-\bar u_\infty(\bar x+\bar b,\tau).
	\label{eqimgbase}
\end{equation}
For the second term of Eq.\ \eqref{eqimg4b}, the product $\cos\left( \kappa (\bar x\mp\bar b)\right)\cos(2j\bar L \kappa )$ is expanded using product-to-sum cosine formula. This gives
\begin{equation}
	\cos\left( \kappa (\bar x-\bar b)\right)\cos(2j\bar L \kappa )=\dfrac{1}{2}\left[\cos\left( \kappa (\bar x-\bar b-2j\bar L)\right)+\cos\left( \kappa (\bar x-\bar b+2j\bar L)\right)\right],
	\label{eqimgshift1}
\end{equation}
and similarly for $\cos\left( \kappa (\bar x+\bar b)\right)\cos(2j\bar L \kappa )$ with $\bar b$ replaced by $-\bar b$ in Eq.~\eqref{eqimgshift1}. Substituting these into the second term of Eq.\ \eqref{eqimg4b} and again using Eq.\ \eqref{eqinfx} gives
\begin{equation}
	\dfrac{4\bar P}{\pi}\int_0^\infty f( \kappa ,\tau)\cos(2j\bar L \kappa )\,{\rm d} \kappa 
	=\bar u_\infty(\bar x-\bar b-2j\bar L,\tau)+\bar u_\infty(\bar x-\bar b+2j\bar L,\tau)
	-\bar u_\infty(\bar x+\bar b-2j\bar L,\tau)-\bar u_\infty(\bar x+\bar b+2j\bar L,\tau).
	\label{eqimgshiftresult}
\end{equation}
Collecting Eqs.\ \eqref{eqimgbase} and \eqref{eqimgshiftresult} over all $j\geq1$, and relabelling the resulting terms by a single integer $m$ running over all of $\mathbb{Z}$ (with $m=0$ recovering the base term), the finite beam response yields
\begin{equation}
	\bar u_F(\bar x,\tau)=\sum_{m=-\infty}^{\infty}
	\left[\bar u_\infty(\bar x-\bar b-2m\bar L,\tau)-\bar u_\infty(\bar x+\bar b-2m\bar L,\tau)\right].
	\label{eqimg5}
\end{equation}
Equation \eqref{eqimg5} expresses the finite beam response as a superposition of infinite beam responses from a periodic array of real sources at $\bar b+2m\bar L$ and antisymmetric image sources at $-\bar b+2m\bar L$, which is the familiar image construction for a simply supported boundary, obtained here directly from the modal series.

For the midpoint-struck beam case, $\bar x=\bar b=\bar L/2$, so that $\bar x-\bar b=0$ and $\bar x+\bar b=\bar L$, and Eq.\ \eqref{eqimg5} reduces to
\begin{equation}
	\bar u_F\left(\dfrac{\bar L}{2},\tau\right)=\sum_{m=-\infty}^{\infty}\bar u_\infty(-2m\bar L,\tau)
	-\sum_{m=-\infty}^{\infty}\bar u_\infty\left((1-2m)\bar L,\tau\right).
	\label{eqimgmid1}
\end{equation}
Since $\bar u_\infty(x,\tau)=\bar u_\infty(-x,\tau)$ (see Eq.\ \eqref{eqinfx}), positive and negative image points at the same distance from the source contribute equally. Grouping each such pair together turns the sum over all positive and negative multiples of $\bar L$ into twice the sum over positive multiples only, plus the single term at the origin, giving
\begin{equation}
	\bar u_F\left(\dfrac{\bar L}{2},\tau\right)=\bar u_\infty(0,\tau)
	+2\sum_{m=1}^{\infty}\bar u_\infty(2m\bar L,\tau)
	-2\sum_{j=1}^{\infty}\bar u_\infty\left((2j-1)\bar L,\tau\right).
	\label{eqimgmid2}
\end{equation}
The even and odd multiple sums in Eq.\ \eqref{eqimgmid2} combine into a single sum over all positive integers $n$, with sign $(-1)^n$ selecting even ($+$) or odd ($-$) multiples of $\bar L$, giving
\begin{equation}
	\bar u_F\left(\dfrac{\bar L}{2},\tau\right)=\bar u_\infty(0,\tau)+2\sum_{j=1}^{\infty}(-1)^j\bar u_\infty(j\bar L,\tau).
	\label{eqimg6}
\end{equation}
This is the finite-length correction relation for the midpoint case. The finite-length correction is then defined as
\begin{equation}
	R_u(\tau,\bar L)=\bar u_F\left(\dfrac{\bar L}{2},\tau\right)-\bar u_\infty(0,\tau)
	=2\sum_{j=1}^{\infty}(-1)^j\bar u_\infty(j\bar L,\tau)=\dfrac{2\bar P}{\pi}\sum_{j=1}^{\infty}(-1)^j\int_0^\infty
	\dfrac{\sin\left(\dfrac{ \kappa ^2}{\sqrt{1+ \kappa ^2}}\tau\right)\cos(j\bar L \kappa )}{ \kappa ^2\sqrt{1+ \kappa ^2}}\,{\rm d} \kappa .
	\label{eqimg7}
\end{equation}
The corresponding velocity correction obtained by differentiating Eq.\ \eqref{eqimg7} with respect to $\tau$,
\begin{equation}
	R_v(\tau,\bar L)=\dfrac{\partial R_u}{\partial\tau}=2\sum_{j=1}^{\infty}(-1)^j\bar v_\infty(j\bar L,\tau)=\dfrac{2\bar P}{\pi}\sum_{j=1}^{\infty}(-1)^j\int_0^\infty
	\dfrac{\cos\left(\dfrac{ \kappa ^2}{\sqrt{1+ \kappa ^2}}\tau\right)\cos(j\bar L \kappa )}{1+ \kappa ^2}\,{\rm d} \kappa .
	\label{eqimg8}
\end{equation}

The integral in Eq.\ \eqref{eqimg7} can be evaluated exactly in Laplace domain. Taking the Laplace transform with respect to $\tau$, and using the same kernel as in Eq.\ \eqref{eqsecn13},
\begin{equation}
	\mathcal{L}\left\{ R_u\right\}=\dfrac{2\bar P}{\pi}\sum_{j=1}^{\infty}(-1)^j\int_0^\infty
	\dfrac{\cos(j\bar L \kappa )}{ \kappa ^4+s^2 \kappa ^2+s^2}\,{\rm d} \kappa .
	\label{eqlap1}
\end{equation}
The quartic denominator factors as $( \kappa ^2+\alpha^2)( \kappa ^2+\beta^2)$, with
\begin{equation}
	\alpha=\sqrt{\dfrac{s}{2}\left(s+\sqrt{s^2-4}\right)}, \qquad
	\beta=\sqrt{\dfrac{s}{2}\left(s-\sqrt{s^2-4}\right)}.
	\label{eqlap2}
\end{equation}
Using partial fractions and the standard cosine integral $\displaystyle \int_0^\infty\cos(c \kappa )/( \kappa ^2+a^2)\,{\rm d} \kappa =\pi e^{-ac}/2a$, each term in Eq.\ \eqref{eqlap1} reduces to an exponential in $j$, and the resulting geometric series, $\displaystyle \sum_{j=1}^\infty(-1)^je^{-jz}=-1/(e^z+1)$, can be summed in closed form. This gives
\begin{equation}
	\mathcal{L}\left\{ R_u\right\}=\dfrac{\bar P}{s\sqrt{s^2-4}}
	\left[\dfrac{1}{\alpha\left(e^{\alpha\bar L}+1\right)}-\dfrac{1}{\beta\left(e^{\beta\bar L}+1\right)}\right],
	\label{eqlap3}
\end{equation}
which is the exact closed-form Laplace-domain expression for the finite-length correction, in terms of $\alpha$ and $\beta$ of Eq.\ \eqref{eqlap2}. Equation~\eqref{eqlap3} provides the short-time asymptotic expansion of the correction factor $R_u$ through a large-$s$ expansion.






\bibliographystyle{unsrt}
\bibliography{cas-refs}

@article{BHATTACHARJEE2018147,
title = {{Transverse impact of a Hertzian body with an infinitely long Euler-Bernoulli beam}},
journal = {Journal of Sound and Vibration},
volume = {429},
pages = {147-161},
year = {2018},
issn = {0022-460X},
author = {Arindam Bhattacharjee and Anindya Chatterjee},
doi = {10.1016/j.jsv.2018.05.020}
}

@article{clarisse1996integrals,
title = {{Integrals with a large parameter: water waves on finite depth due to an impulse}},
author = {J-M Clarisse and J. N. Newman and F. Ursell},
journal = {Oceanographic Literature Review},
volume = {2},
number = {43},
pages = {102},
year = {1996}
}

@incollection{serov2017riemann,
title = {{The Riemann--Lebesgue lemma}},
author = {Valery Serov},
booktitle = {Fourier Series, Fourier Transform and Their Applications to Mathematical Physics},
pages = {33-35},
year = {2017},
address = {Cham},
publisher = {Springer}
}

@book{whitham2011linear,
title = {{Linear and Nonlinear Waves}},
author = {Gerald Beresford Whitham},
year = {2011},
address = {New York},
publisher = {John Wiley \& Sons},
note = {Originally published in 1974}
}

@book{timoshenko1930strength,
title = {{Strength of Materials}},
author = {Stephen Timoshenko},
year = {1930},
address = {New York},
publisher = {D. Van Nostrand Company}
}

@book{graff2012wave,
title = {{Wave Motion in Elastic Solids}},
author = {Karl F. Graff},
year = {1991},
address = {New York},
publisher = {Dover Publications},
note = {Originally published in 1975}
}

@article{goswami2023short,
title = {{Short time angular impulse response of Rayleigh beams}},
author = {Bidhayak Goswami and K. R. Jayaprakash and Anindya Chatterjee},
journal = {Journal of Engineering Mathematics},
volume = {143},
number = {1},
pages = {6},
year = {2023},
publisher = {Springer},
doi = {10.1007/s10665-023-10300-8}
}

@book{meirovitch1967analytical,
title = {{Analytical Methods in Vibrations}},
author = {Leonard Meirovitch},
year = {1967},
address = {New York},
publisher = {The Macmillan Company}
}

@article{chatterjee2004short,
title = {{The short-time impulse response of Euler-Bernoulli beams}},
author = {Anindya Chatterjee},
journal = {Journal of Applied Mechanics},
volume = {71},
number = {2},
pages = {208-218},
year = {2004},
doi = {10.1115/1.1640369}
}

@book{rayleigh1894theory,
title = {{The Theory of Sound}},
author = {John William Strutt},
volume = {1},
series = {Cambridge Library Collection -- Physical Sciences},
address = {Cambridge},
publisher = {Cambridge University Press},
year = {2011},
note = {Originally published in 1894}
}

@article{timoshenko1921,
title = {{On the correction for shear of the differential equation for transverse vibrations of prismatic bars}},
author = {Stephen P. Timoshenko},
journal = {Philosophical Magazine, Series 6},
volume = {41},
number = {245},
pages = {744-746},
year = {1921}
}

@article{han1999dynamics,
title = {{Dynamics of transversely vibrating beams using four engineering theories}},
author = {S. M. Han and H. Benaroya and T. Wei},
journal = {Journal of Sound and Vibration},
volume = {225},
number = {5},
pages = {935-988},
year = {1999},
doi = {10.1006/jsvi.1999.2257}
}

@article{cowper1966shear,
title = {{The shear coefficient in Timoshenko's beam theory}},
author = {G. R. Cowper},
journal = {Journal of Applied Mechanics},
volume = {33},
number = {2},
pages = {335-340},
year = {1966},
doi = {10.1115/1.3625046}
}

@article{traillnash1953effects,
title = {{The effects of shear flexibility and rotatory inertia on the bending vibrations of beams}},
author = {R. W. Traill-Nash and A. R. Collar},
journal = {Quarterly Journal of Mechanics and Applied Mathematics},
volume = {6},
number = {2},
pages = {186-222},
year = {1953},
doi = {10.1093/qjmam/6.2.186}
}

@article{zak2012assessment,
title = {Assessment of flexural beam behaviour theories used for dynamics and wave propagation problems},
journal = {Journal of Sound and Vibration},
volume = {331},
number = {26},
pages = {5715-5731},
year = {2012},
issn = {0022-460X},
doi = {https://doi.org/10.1016/j.jsv.2012.07.034},
url = {https://www.sciencedirect.com/science/article/pii/S0022460X12005603},
author = {A. Żak and M. Krawczuk}
}

@article{stephen2006second,
title = {The second spectrum of Timoshenko beam theory—Further assessment},
journal = {Journal of Sound and Vibration},
volume = {292},
number = {1},
pages = {372-389},
year = {2006},
issn = {0022-460X},
doi = {https://doi.org/10.1016/j.jsv.2005.08.003},
url = {https://www.sciencedirect.com/science/article/pii/S0022460X05005213},
author = {N.G. Stephen}
}

@book{brillouin1960wave,
title = {{Wave Propagation and Group Velocity}},
author = {L. Brillouin},
publisher = {Academic Press},
address = {New York},
year = {1960}
}

@article{dolph1954,
  title={{On the Timoshenko theory of transverse beam vibrations}},
  author={Dolph, CL},
  journal={Quarterly of Applied Mathematics},
  volume={12},
  number={2},
  pages={175--187},
  year={1954}
}

@article{herrmann1955,
    author = {Herrmann, G.},
    title = {{Forced Motions of Timoshenko Beams}},
    journal = {Journal of Applied Mechanics},
    volume = {22},
    number = {1},
    pages = {53-56},
    year = {2021},
    month = {06},
    issn = {0021-8936},
    doi = {10.1115/1.4010969},
    url = {https://doi.org/10.1115/1.4010969},
    eprint = {https://asmedigitalcollection.asme.org/appliedmechanics/article-pdf/22/1/53/6748941/53_1.pdf},
}

@article{levinson1982,
title = {{On the two frequency spectra of Timoshenko beams}},
journal = {Journal of Sound and Vibration},
volume = {84},
number = {3},
pages = {319-326},
year = {1982},
issn = {0022-460X},
doi = {https://doi.org/10.1016/0022-460X(82)90480-1},
url = {https://www.sciencedirect.com/science/article/pii/0022460X82904801},
author = {M. Levinson and D.W. Cooke}
}

\end{document}